\documentclass{LMCS}

\def\doi{8 (1:07) 2012}
\lmcsheading%
{\doi}
{1--26}
{}
{}
{Dec.~30, 2010}
{Feb.~20, 2012}
{}

\usepackage{enumerate}
\usepackage{hyperref}
\usepackage{amssymb}

\newcommand{\indg}[2]{#1_{|#2}}
\newcommand{\graph}[1]{\mathbb{#1}}
\newcommand{\absorb}{\triangleleft }
\newcommand{\minabs}{\triangleleft\!\triangleleft \ }
\newcommand{\comp}{*}

\newcommand{\mdiv}{\ \mathrm{div} \ }
\newcommand{\mmod}{\ \mathrm{mod} \ }

\newcommand{\arity}[1]{\mathrm{ar}(#1)}
\newcommand{\tuple}[1]{\mathbf{#1}}
\newcommand{\fence}[2]{{\mathbb F}[#1,#2]}
\newcommand{\Fence}{{\mathbb F}}
\newcommand{\Fencepr}{{\mathbb F'}}

\newcommand{\Clo}{\mathrm{Clo}}
\newcommand{\CSP}{\mathrm{CSP}}
\newcommand{\Idpol}{\mathrm{IdPol}}

\newcommand{\variety}[1]{\mathcal{#1}}
\newcommand{\relstr}[1]{\mathbb{#1}}
\newcommand{\alg}[1]{\mathbf{#1}}

\newcommand{\Sg}[2]{\mathrm{Sg}_{#1}(#2)}

\theoremstyle{plain}
\newtheorem*{fedvardich}{The dichotomy conjecture of Feder and Vardi}
\newtheorem*{algdich}{The Algebraic Dichotomy Conjecture}
\theoremstyle{remark}
\newtheorem{claim}[thm]{Claim}

\begin{document}

\title[Absorption, cyclic terms, and CSP]{Absorbing Subalgebras, Cyclic Terms, and the Constraint Satisfaction Problem\rsuper*}

\author[L.~Barto]{Libor Barto\rsuper a}
\address{{\lsuper a}Department of Algebra, Charles University, Sokolovsk\'a 83, 186 75 Prague 8, Czech Republic \and Department of Mathematics and Statistics, McMaster University, 1280 Main Street West, Hamilton, ON, L8S 4K1, Canada}
\email{libor.barto@gmail.com} 
\thanks{{\lsuper a}Research supported by the Grant Agency of the Czech Republic under the grant No. 201/09/P223 and 
by the Ministry of Education of the Czech Republic under the grant 
No. MSM 0021620839.}

\author[M.~Kozik]{Marcin Kozik\rsuper b}	
\address{{\lsuper b}Theoretical Computer Science Department, Faculty
  of Mathematics and Computer Science, Jagiellonian University,
ul. Prof. St. \L ojasiewicza 6, 30-348 Krak\'ow, Poland}	
\email{marcin.kozik@tcs.uj.edu.pl}  \
\thanks{{\lsuper b}Research supported by the Foundation for Polish Science under the grant No. HOM/2008/7 (supported by MF EOG), 
and Ministry of Science and Higher Education of Poland under the grant No. N206 357036.}

\titlecomment{{\lsuper*}A part of this work has appeared in our paper \emph{New conditions for Taylor varieties and CSP}, Proceedings of the 25th IEEE Symposium on Logic in Computer Science, LICS'10, 100-109}

\begin{abstract}
  \noindent
The Algebraic Dichotomy Conjecture states that the Constraint
Satisfaction Problem over a fixed template is solvable in polynomial
time if the algebra of polymorphisms associated to the template lies
in a Taylor variety, and is NP-complete otherwise.
This paper provides two new characterizations of finitely generated
Taylor varieties. The first characterization is using
absorbing subalgebras and the second one cyclic terms. These new
conditions allow us to reprove the conjecture of  Bang-Jensen and Hell
(proved by the authors)
and the characterization of locally finite Taylor varieties using weak
near-unanimity terms  (proved by McKenzie and Mar\' oti) in an elementary and self-contained way. 

\end{abstract}

\maketitle

\section*{Introduction}

The Constraint Satisfaction Problem (CSP) is a generic problem in computer science. An instance consists of a number of variables and constraints imposed on them and the objective is to determine whether variables can be assigned values in such a way that all the constraints are met.
As CSP provides a common framework for many theoretical problems as well as for many real-life applications, it has been studied by computer scientists for over forty years.

The results contained in this paper follow a long line of research
devoted to verifying the Constraint Satisfaction Problem Dichotomy
Conjecture of Feder and Vardi~\cite{FV99}. It deals with so called
\emph{non-uniform} CSP --- 
the same decision problem as the ordinary CSP, but in this case the set of allowed constraint relations
is finite and fixed. The conjecture states that, for every finite,
fixed set of constraint relations~(a fixed \emph{template}), the CSP defined by it is NP-complete or solvable in polynomial time, i.e. the class of CSPs exhibits a dichotomy.

The conjecture of Feder and Vardi dates back to 1993. At that time it was supported by two major results, Schaefer's dichotomy theorem for two--element templates \cite{shaefer2}, and the dichotomy theorem for undirected graphs by Hell and Ne\v set\v ril \cite{HN90}. 
The first breakthrough in the research appeared in 1997 in the work of Jeavons, Cohen and Gyssens~\cite{JCG97}, refined later by Bulatov, Jeavons and Krokhin~\cite{BKJ00, BJK05}. 
At heart of the new approach lies a proof that the complexity of CSP, for a fixed template, depends only on a set of certain operations --- polymorphisms of the template. Thus the study of templates gives rise to the study of algebras associated to them. 

The algebraic approach has lead to a better understanding of the known
results and brought a number of new results which were out of reach
for pre-algebraic methods. The theorem of Schaefer~\cite{shaefer2} has
been extended by Bulatov~\cite{B06} to three--element domains. 
Another major result of Bulatov~\cite{B03,Bul11} establishes the dichotomy
for templates containing all unary relations.
The conjecture of Bang-Jensen and Hell~\cite{BJH}, generalizing Hell's
and Ne\v set\v ril's dichotomy theorem \cite{HN90}, was confirmed~\cite{smoothproc,smoothpaper}. 
New algorithms were devised~\cite{BD06,D06,FewSPLICS} and 
pre-algebraic algorithms were characterized in algebraic terms~\cite{cdbw,bwproc}.

The hardness parts in the dichotomy results mentioned above were
obtained using a theorem of Bulatov, Jeavons and
Krokhin~\cite{BKJ00,BJK05} stating that whenever an algebra associated
with a core template does not lie in a Taylor variety then the CSP
defined by the template is NP-complete. 
In the same paper the authors conjecture that in all the other cases
the associated CSP is solvable in polynomial time.  
All the known partial results agree with this proposed classification,
which is now commonly referred to as the Algebraic Dichotomy
Conjecture.

In order to prove the Algebraic Dichotomy Conjecture one has to devise
an algorithm that works for any relational structure with the
corresponding algebra in a Taylor variety. 
As the characterization originally provided by Taylor~\cite{taylor} is difficult to
work with, a search for equivalent
conditions is ongoing. 
A technical, but useful condition was obtained by Bulatov who used it
to prove his dichotomy theorems~\cite{B03,B06}.
Another powerful tool is the characterization of (locally finite)
Taylor
varieties in terms of weak near-unanimity operations due to Mar\' oti and
McKenzie~\cite{MM}.
Unfortunately, their proof uses a deep algebraic theory of Hobby and
McKenzie~\cite{TCTbook}, therefore is not easily accessible for a
nonspecialist. The proof of the conjecture of
Bang-Jensen and Hell hinges on this characterization; also the
algebraic characterization of problems of bounded width~\cite{bwproc}
relies on a similar characterization of congruence meet
semi-distributive varieties provided in the same paper~\cite{MM}.
Recently, a surprisingly simple condition for Taylor varieties was found by
Siggers~\cite{sig}, and an analytical characterization was given by
Kun and Szegedy~\cite{KS09}.

In this paper we provide two new conditions for (finitely
generated) Taylor varieties. These new characterizations already
proved to be useful. Not only they provide new tools for attacking the
algebraic dichotomy conjecture, but they also allow us to present easy and
elementary proofs for some of the results mentioned above. 
Moreover, their proofs  are self-contained
and do not require heavy algebraic machinery.

The first, structural characterization (the Absorption Theorem) is 
expressed in terms of \emph{absorbing subalgebras}
developed and successfully applied by the authors
in~\cite{smoothproc,smoothpaper,cdbw,bwproc}. We use it to present an
elementary proof of the conjecture of Bang-Jensen and Hell. Recently,
the Absorption Theorem was applied to give a short proof of Bulatov's
dichotomy theorem for conservative CSPs~\cite{conserv}.
The second, equational characterization involves \emph{cyclic terms} and is a stronger version of
the weak near-unanimity condition. 
We use it to restate
the Algebraic Dichotomy Conjecture in simple combinatorial terms
and to provide a very short proof of the theorem of Hell and Ne\v set\v ril.

The results of this paper also show that the tools developed for the
CSP can be successfully applied to algebraic questions which indicates
a deep connection between the CSP and universal algebra.

\subsection*{Organization of the paper}
In section~\ref{sect:preliminaries} we introduce the necessary notions
concerning algebras and the CSP. In section~\ref{sect:absorbing} we
define absorbing subalgebras and present the Absorption Theorem and
its corollaries. In section~\ref{sect:smoothproof} we use the
absorbing subalgebra characterization to provide an elementary proof
of the conjecture of Bang-Jensen and Hell in a slightly stronger
version which is needed in section~\ref{sect:cyclic}. 
Finally, in section~\ref{sect:cyclic} we prove the characterization using cyclic terms and its corollaries: the theorem of Hell and Ne\v set\v ril~\cite{HN90} and the weak near-unanimity characterization of locally finite Taylor varieties of Mar\'oti and McKenzie~\cite{MM}.

\section{Preliminaries}\label{sect:preliminaries}

\subsection{Notation for sets}
\noindent For a set $A$ and a natural number $n$, elements of $A^n$ are the $n$-tuples of elements of $A$. We index its coordinates starting from zero, for example $(a_0, a_1, \dots, a_{n-1}) \in A^n$.

Let $R$ be a subset of a Cartesian product $A_1 \times A_2 \times \dots \times A_n$. $R$ is called \emph{subdirect}~($R \subseteq_S A_1 \times \dots \times A_n$) if, for every $i=1,2,\dotsc,n$,  the projection of $R$ to the $i$-th coordinate is the whole set $A_i$.

Given $R \subseteq A \times B$ and $S \subseteq B \times C$, by $S \circ R$ we mean the following subset of $A \times C$:
\begin{equation*}
S \circ R = \{ (a,c) : \exists \ b \in B \ \ (a,b) \in R, (b,c) \in S \}.
\end{equation*}
If $R \subseteq A \times A$ and $n$ is a natural number greater than zero, then we define
$$
R^{{} \circ n} = \underbrace{R \circ R \circ \dots \circ R}_{n}.
$$

\subsection{Algebras and varieties}\label{sect:algbasics}

\noindent An \emph{algebraic signature} is a finite set of function symbols with a natural number~(the \emph{arity}) associated to each of them. An \emph{algebra} of 
a signature $\Sigma$ is a pair $\alg{A} = (A, (t^{\alg{A}})_{t \in \Sigma})$, where $A$ is a set, called the \emph{universe} of $\alg{A}$, and $t^{\alg{A}}$ is an operation on $A$ of arity $\arity{t}$, that is, a mapping $A^{\arity{t}} \rightarrow A$. 
We always use a boldface letter to denote an algebra and the same letter in a plain type to denote its universe. We often omit the superscripts of operations when the algebra is clear from the context.

A \emph{term} in a signature $\Sigma$ is a formal expression using variables and compositions of symbols in $\Sigma$. In this paper we introduce a special notation for a particular case of composition of terms: given a $k$-ary term $t_1$ and an $l$-ary term $t_2$ we define 
a $kl$-ary term $t_1 \comp t_2$ by
\begin{equation*}
t_1 \comp t_2 (x_0, x_1, \dots, x_{kl-1}) 
=
t_1(t_2(x_0,  \dots, x_{l-1}), t_2(x_l, \dots,x_{2l-1}), \dots, t_2(x_{(k-1)l} \dots, x_{kl-1})).
\end{equation*}
For an algebra $\alg{A}$ and a term $h$ in the same signature $\Sigma$, $h^{\alg{A}}$ has the natural meaning in $\alg{A}$ and is called a \emph{term operation} of $\alg{A}$. Again, we usually omit the superscripts of term operations when the algebra is clear from the context. The set of all term operations of $\alg{A}$ is called the \emph{clone of term operations} of $\alg{A}$ and it is denoted $\Clo(\alg{A})$. 

For a pair of terms $s,t$ over a signature $\Sigma$, we say that an algebra $\alg{A}$ in the signature $\Sigma$ \emph{satisfies the
identity} $s \approx t$ if the term operations $s^{\alg{A}}$ and $t^{\alg{A}}$ are the same.

There are three fundamental operations on algebras of a fixed signature $\Sigma$: forming  subalgebras, factoralgebras and products. A subset $B$ of the universe of an algebra $\alg{A}$ is called a \emph{subuniverse}, if it is closed under all operations (equivalently term operations) of $\alg{A}$. Given a subuniverse $B$ of $\alg{A}$ we can form the algebra $\alg{B}$ by restricting all the operations of $\alg{A}$ to the set $B$. In this situation we write $B \leq \alg{A}$ or $\alg{B} \leq \alg{A}$. We call the subuniverse $B$ (or the subalgebra $\alg{B}$) \emph{proper} if $\emptyset \neq B \neq A$. The smallest subalgebra of $\alg{A}$ containing a set $B \subseteq A$ is called the subalgebra \emph{generated by} $B$ and will be denoted by $\Sg{\alg{A}}{B}$. It can be equivalently described as  the set of elements which can be obtained by applying term operations of $\alg{A}$ to elements of $B$. 

Given a family of algebras $\alg{A}_i, i \in I$ we define its product $\prod_{i \in I} \alg{A}_i$ to be the algebra with the universe equal to the cartesian
product of the $A_i$'s and with operations computed coordinatewise. The product of algebras $\alg{A}_1$, \dots, $\alg{A}_n$ will be denoted by
$\alg{A}_1 \times \dots \times \alg{A}_n$ and the product of $n$ copies of an algebra $\alg{A}$ by $\alg{A}^n$. 
$\alg{R}$ is a \emph{subdirect subalgebra} of $\alg{A}_1 \times \alg{A}_2 \times \dotsb \times \alg{A}_n$ if $R$ is subdirect in $A_1\times A_2\times \dotsb \times A_n$ and, in such a case, we write $\alg{R} \leq_S \alg{A}_1 \times \dots \times \alg{A}_n$.

An equivalence relation $\sim$ on the universe of an algebra $\alg{A}$ is a \emph{congruence}, if it is a subalgebra of $\alg{A}^2$. The corresponding \emph{factor algebra} $\alg{A}/\!\sim$ has, as the universe, the set of $\sim$-blocks and the operations are defined using (arbitrarily chosen) representatives. A congruence is \emph{nontrivial}, if it is not equal to the diagonal or to the full relation $A \times A$.

A \emph{variety} is a class of algebras of the same signature closed under forming isomorphic copies, subalgebras, factoralgebras and products. For a pair of terms $s,t$ over a signature $\Sigma$, we say that a class of algebras $\variety{V}$ in the signature $\Sigma$ \emph{satisfies the
identity $s \approx t$} if every algebra in the class does. By Birkhoff's theorem, a class of algebras is a variety if and only if there exists a set of identities $E$ such that the members of $\variety{V}$ are precisely those algebras which satisfy all the identities from $E$. 

A variety $\variety{V}$ is called \emph{locally finite}, if every finitely generated algebra (that is, an algebra generated by a finite subset) contained in $\variety{V}$  is finite.  $\variety{V}$ is called \emph{finitely generated}, if there exists a finite set $\variety{K}$ of finite algebras such that $\variety{V}$ is the smallest variety containing $\variety{K}$. In such a case $\variety{V}$ is actually generated by a single, finite algebra, the product of members of $\variety{K}$. Every finitely generated variety is locally finite, and if a variety is generated by a single algebra then the identities satisfied in this algebra are exactly the identities satisfied in the variety.

For a more in depth introduction to universal algebra and proofs of the above mentioned results we recommend~\cite{BS81}.

\subsection{Taylor varieties}
\noindent A term $s$ is \emph{idempotent} in a variety~(or an algebra), if it satisfies the identity
\begin{equation*}
s(x,x, \dots, x) \approx x.
\end{equation*}
An algebra~(a variety) is idempotent if all its terms are.

A term $t$ of arity at least $2$ is called a \emph{weak near-unanimity} term of a variety~(or an algebra), if $t$ is idempotent and satisfies
\begin{equation*}
t(y,x,x, \dots, x) \approx t(x,y,x,x, \dots, x)\approx\dots
  \dots \approx t(x,x, \dots, y,x)\approx t(x,x, \dots, x,y).
\end{equation*}
A term $t$ of arity at least $2$ is called a \emph{cyclic} term of a variety~(or an algebra), if $t$ is idempotent and satisfies
\begin{equation*}
t(x_0, x_1, \dots, x_{k-1}) \approx t(x_1, x_2, \dots, x_{k-1}, x_0).
\end{equation*}
Finally, a 
term $t$ of arity $k$ is called a \emph{Taylor term} of a variety~(or an algebra), if $t$ is idempotent and for every $j<k$ it satisfies an identity of the form 
\begin{equation*}
t(\Box_0, \Box_1, \dots, \Box_{k-1}) \approx t(\triangle_0, \triangle_1, \dots, \triangle_{k-1}), 
\end{equation*}
where all $\Box_i$'s and $\triangle_i$'s are substituted with either $x$ or $y$, but $\Box_j$ is $x$ while $\triangle_j$ is $y$.

\begin{defi}
An idempotent variety $\variety{V}$ is called \emph{Taylor} if it has a Taylor term.
\end{defi}

\noindent
Study of Taylor varieties has been a recurring subject in universal algebra for many years. One of the first characterizations is due to Taylor~\cite{taylor}
\begin{thm}[Taylor~\cite{taylor}] \label{thm:taylor}
Let $\variety{V}$ be an idempotent variety. The following are equivalent.
\begin{iteMize}{$\bullet$}
\item $\variety{V}$ is a Taylor variety.
\item $\variety{V}$ does not contain a two-element algebra whose every (term) operation is a projection.
\end{iteMize}
\end{thm}

\noindent Further research led to discovery of other equivalent conditions~\cite{TCTbook, MM, sig, KS09}. 
One of the most important ones is the result of Mar\'oti and McKenzie~\cite{MM}.
\begin{thm}[Mar\'oti and McKenzie~\cite{MM}] \label{thm:wnu}
Let $\variety{V}$ be an idempotent, locally finite variety. The following are equivalent.
\begin{iteMize}{$\bullet$}
\item $\variety{V}$ is a Taylor variety.
\item $\variety{V}$ has a weak near-unanimity term.
\end{iteMize}
\end{thm}
\noindent This result, together with a similar characterization provided in the same paper for congruence meet semi-distributive varieties, found deep applications in CSP~\cite{smoothproc, smoothpaper, bwproc}.

\subsection{Relational structures and CSP}

A convenient formalization of non-uniform CSP is via homomorphisms between relational structures~\cite{FV99}.

\noindent A \emph{relational signature} is a finite set of relation symbols with arities associated to them. 
A \emph{relational structure} of 
the signature $\Sigma$ is a pair $\relstr{A} = (A, (R^{\relstr{A}})_{R \in \Sigma})$, where $A$ is a set, called the \emph{universe} of $\relstr{A}$, and $R^{\relstr{A}}$ is a relation  on $A$ of arity $\arity{R}$, that is, a subset of $A^{\arity{R}}$. 

Let $\relstr{A},\relstr{B}$ be relational structures of the same signature. A mapping $f: A \rightarrow B$ is a \emph{homomorphism} from $\relstr{A}$ to $\relstr{B}$, if
it preserves all $R \in \Sigma$, that is, $(f(a_0), f(a_1), \dots, f(a_{\arity{R}-1})) \in R^{\relstr{B}}$ for any $(a_0, \dots, a_{\arity{R}-1}) \in R^{\relstr{A}}$. A finite relational structure $\relstr{A}$ is a \emph{core}, if every homomorphism from $\relstr{A}$ to itself is bijective.

For a fixed relational structure $\relstr{A}$ of a signature $\Sigma$, $\CSP(\relstr{A})$ is the following decision problem:

\begin{tabular}{ll}
INPUT: & A relational structure $\relstr{X}$ of the signature $\Sigma$. \\
QUESTION: & Does $\relstr{X}$ map homomorphically to $\relstr{A}$?
\end{tabular}

\noindent It is easy to see that if $\relstr{A}'$ is a core of $\relstr{A}$~(i.e. a core which is contained in $\relstr{A}$ and such that $\relstr{A}$ can be mapped homomorphically into it) then $\CSP(\relstr{A})$ and $\CSP(\relstr{A}')$ are identical.

The celebrated conjecture of Feder and Vardi~\cite{FV99} states that the class of $\CSP$s exhibits a dichotomy:
\begin{fedvardich}
 For any relational structure $\relstr{A}$, the problem $\CSP(\relstr{A})$ is solvable in polynomial time, or NP-complete.
\end{fedvardich}

\subsection{Algebraic approach to CSP}
\noindent A mapping $f: A^n \rightarrow A$ is \emph{compatible} with an $m$-ary relation $R$ on $A$ if the tuple
\begin{equation*}
\big(f(a^0_0, a^1_0, \dots, a^{n-1}_0),
\dots, 
f(a^0_{m-1}, a^1_{m-1}, \dots, a^{n-1}_{m-1})\big)
\end{equation*}
belongs to $R$ whenever $(a^i_0,\dotsc,a^i_{m-1}) \in R$ for all $i<n$. A mapping compatible with all the relations in a relational structure $\relstr{A}$ is a \emph{polymorphism} of this structure.

For a given relational structure $\relstr{A} = (A, (R^{\relstr{A}})_{R\in\Sigma})$ we define an algebra $\Idpol(A)$~(often denoted by just $\alg{A}$). This algebra $\alg{A}$ has its universe equal to $A$ and the operations of $\alg{A}$ are the idempotent polymorphisms of $\relstr{A}$~(we formally define a signature of $\alg{A}$ to be identical with the set of its operations).

It follows from an old result~\cite{Dual1, Dual2} that a relation $R$ of arity $k$ is a subuniverse of $\Idpol(\relstr{A})^k$ if and only if $R$ can be positively primitively defined from relations in $\relstr{A}$ and singleton unary relations identifying every element of $A$. That is, $R$ can be defined by a first-order formula which uses relations in $\relstr{A}$, singleton unary relations on $A$, the equality relation on $A$, conjunction and existential quantification. 

Already the first results on the algebraic approach to
CSP~\cite{JCG97,BKJ00,BJK05} show that whenever a relational structure
$\relstr{A}$ is a core then $\Idpol(\relstr{A})$ fully determines the
computational complexity of $\CSP(\relstr{A})$. Moreover, Bulatov,
Jeavons and Krokhin showed~\cite{BKJ00, BJK05}:

\begin{thm}[Bulatov, Jeavons and Krokhin~\cite{BKJ00,BJK05}] \label{thm:bjk}
Let $\relstr{A}$ be a finite relational structure which is a core. If $\ \Idpol(\relstr{A})$ does not lie in a Taylor variety, then $\CSP(\relstr{A})$ is $NP$-complete.
\end{thm}

\noindent In the same paper they conjectured that these are the only cases of finite cores which give rise to NP-complete CSPs.
\begin{algdich}
Let $\relstr{A}$ be a finite relational structure which is a core. If $\ \Idpol(\relstr{A})$ does not lie in a Taylor variety, then $\CSP(\relstr{A})$ is $NP$-complete. Otherwise is it solvable in polynomial time.
\end{algdich}

\noindent
This conjecture is supported by many partial results on the complexity of CSPs~\cite{B03, B06, smoothproc, smoothpaper, bwproc, FewSPLICS} and it renewed interest in properties of finitely generated Taylor varieties.

\section{Absorbing subalgebras and absorption theorem}\label{sect:absorbing}

\noindent In this section we introduce the concept of an absorbing subalgebra and prove the Absorption Theorem and its corollaries.
The proof is self-contained and elementary. In section~\ref{sect:smoothproof} we use Theorem~\ref{thm:abs} to reprove a stronger version of the ``Smooth Theorem''~\cite{smoothproc,smoothpaper} which, in turn, will be used to prove the second main result of this article, Theorem~\ref{thm:cyclic}.  This approach simplifies significantly the known proof of the Smooth Theorem, and does not rely on the involved algebraic results results from~\cite{MM}.
It has also lead to a simple proof \cite{conserv} of the dichotomy theorem for conservative CSPs \cite{B03}.

\subsection{Absorption}
A subalgebra $B$ of an algebra $\alg{A}$ is an absorbing subalgebra, if there exists a term operation of $\alg{A}$ which outputs an element of $B$ whenever all but at most one of its arguments are from $B$. More precisely

\begin{defi}
Let $\alg{A}$ be an algebra and $t \in \Clo(\alg{A})$. We say that a subalgebra $\alg{B}$ of $\alg{A}$ is an \emph{absorbing subalgebra of $\alg{A}$ with respect to $t$} if, for any $k < \arity{t}$ and any choice of $a_i \in A$ such that $a_i\in B$ for all $i\neq k$,  we have $t(a_0,\dotsc, a_{\arity{t}-1}) \in B$.

We say that $\alg{B}$ is an \emph{absorbing subalgebra of $\alg{A}$}, or that \emph{$\alg{B}$ absorbs $\alg{A}$}~(and write $\alg{B} \absorb \alg{A}$), if there exists $t \in \Clo(\alg{A})$ such that $\alg{B}$ is an absorbing subalgebra of $\alg{A}$ with respect to $t$. 
\end{defi}

\noindent
We also speak about \emph{absorbing subuniverses}, i.e. universes of absorbing subalgebras. 
Recall that an (absorbing) subalgebra $\alg{B}$ of $\alg{A}$ is \emph{proper}, if $\emptyset \neq B \varsubsetneq A$.

The Absorption Theorem says that the existence of a certain kind of subuniverse $R$ of a product of two Taylor algebras $\alg{A}$ and $\alg{B}$ forces a proper absorbing subuniverse in one of these algebras. 
It is helpful to draw $R$ as a bipartite undirected graph in the following sense: the vertex set is the disjoint union of $A$ (draw it on the left) and $B$ (on the right) and two elements $a \in A$ from the left side and  $b \in B$ from the right side are adjacent  if $(a,b) \in R$.  
We say that two vertices are linked if they are connected in this graph, and 
we call $R$ linked if the graph is connected after deleting the isolated vertices. Note that $R \leq_S \alg{A} \times \alg{B}$ if and only if there are no isolated vertices.

\begin{defi}
 Let $R \subseteq A \times B$ and let $a,a' \in A$. We say that $a,a' \in A$ are \emph{linked  in $R$}, or \emph{$R$-linked}, via $c_0, \dots, c_{2n}$, if 
$a=c_0, c_{2n}=a'$ and $(c_{2i},c_{2i+1})\in R$ and $(c_{2i+2},c_{2i+1}) \in R$ for all $i=0,1, \dots, n-1$. 

In a similar way we define when $a \in A, a' \in B$ (or $a \in B, a' \in A$, or $a \in B, a' \in B$) are $R$-linked.

We say that $R$  is \emph{linked}, if $a,a'$ are $R$-linked for any elements $a,a'$ of the projection of $R$ to the first coordinate.
\end{defi}

\noindent
These definitions allow us to state the Absorption Theorem which is the first main result of the paper.
\begin{thm}\label{thm:abs}
 Let $\variety{V}$ be an idempotent, locally finite variety, then the following are equivalent.
\begin{iteMize}{$\bullet$}
 \item $\variety{V}$ is a Taylor variety;
 \item for any finite $\alg{A},\alg{B}\in\variety{V}$ and any linked $\alg{R}\leq_S \alg{A}\times\alg{B}$:
\begin{iteMize}{$-$}
\item $\alg{R} = \alg{A}\times\alg{B}$ or
\item $\alg{A}$ has a proper absorbing subuniverse or
\item $\alg{B}$ has a proper absorbing subuniverse.
\end{iteMize}
\end{iteMize}
\end{thm}

\subsection{Proof of Absorption Theorem}
We start with a couple of useful observations. The first one says that 
absorbing subalgebras are closed under taking intersection, and that $\absorb$ is a transitive relation:

\begin{prop} \label{prop:trans}
Let $\alg{A}$ be an algebra. 
\begin{iteMize}{$\bullet$}
\item If $\alg{C} \absorb \alg{B} \absorb \alg{A}$, then $\alg{C} \absorb \alg{A}$.
\item If $\alg{B} \absorb \alg{A}$ and $\alg{C} \absorb \alg{A}$, then $B \cap C \absorb \alg{A}$.
\end{iteMize}
\end{prop}

\proof
We start with a proof of the first item. Assume that $\alg{B}$ absorbs $\alg{A}$ with respect to $t$~(of arity $m$) and that $\alg{C}$ absorbs $\alg{B}$ with respect to $s$~(of arity $n$). We will show that $\alg{C}$ is an absorbing subalgebra of $\alg{A}$ with respect to $s \comp t$. Indeed, take any tuple $(a_0,\dotsc,a_{mn-1}) \in A^{mn}$ such that $a_i \in C$ for all but one index, say $j$, and consider the evaluation of $s\comp t(a_0,\dotsc,a_{mn-1})$. Every evaluation of the term $t$ appearing in $s\comp t$ is of the form 
\begin{equation*}
 t(a_{im},\dotsc,a_{im+m-1})
\end{equation*}
and therefore whenever $j$ does not fall into the interval $[im,im+m-1]$ the result of it falls in $C$~(as $C$ is a subuniverse of $\alg{A}$). In the case when $j$ is in that interval we have a term $t$ evaluated on the elements of $C$~(and therefore elements of $B$) in all except one coordinate. The result of such an evaluation falls in $B$~(as $\alg{B}$ absorbs $\alg{A}$ with respect to $t$). Thus $s$ is applied to a tuple consisting of elements of $C$ on all but one position, and on this position the argument comes from $B$. Since $\alg{C}$ absorbs $\alg{B}$ with respect to $s$ the results falls in $C$ and the first part of the proposition is proved.

For the second part we consider $\alg{B} \absorb \alg{A}$ and $\alg{C} \absorb \alg{A}$; it follows easily that $B \cap C \absorb \alg{C}$ with respect to the same term as $\alg{B} \absorb \alg{A}$. Now it is enough to apply the first part.
\qed

\noindent
Let $R$ be a subuniverse of $\alg{A} \times \alg{B}$. We use the following notation for the neighborhoods of $X \subseteq A$ or $Y \subseteq B$:
\begin{eqnarray*}
X^{+R} &=& \{b \in B: \exists \ a \in X \ \ (a,b) \in R\} \\
Y^{-R} &=& \{a \in A: \exists \ b \in Y \ \ (a,b) \in R\} 
\end{eqnarray*}
When $R$ is clear from the context we write just $X^+$ and $Y^-$. The next lemma shows that these operations preserve~(absorbing) subalgebras.

\begin{lem} \label{lem:neig}
Let $R \leq \alg{A} \times \alg{B}$, where $\alg{A}, \alg{B}$ are algebras of the same signature. If $X \leq \alg{A}$ and $Y \leq \alg{B}$, then
$X^+ \leq \alg{B}$ and $Y^- \leq \alg{A}$. Moreover, if $R \leq_S \alg{A} \times \alg{B}$ and $X \absorb \alg{A}$ and $Y \absorb \alg{B}$, then
$X^+ \absorb \alg{B}$ and $Y^- \absorb \alg{A}$.
\end{lem} 

\proof
Suppose $X \leq \alg{A}$ and take any term $t$, say of arity $j$, in the given signature. Let $b_0,\dotsc,b_{j-1} \in X^+$ be arbitrary.
From the definition of $X^+$ we can find $a_0,\dotsc,a_{j-1} \in X$ such that $(a_i,b_i) \in R$ for all $0 \leq i < j$. 
Since $R$ is a subuniverse of $\alg{A} \times \alg{B}$, the pair $(t(a_0,\dotsc,a_{j-1}), t(b_0,\dotsc,b_{j-1}))$ is in $R$. But $t(a_0,\dotsc,a_{j-1}) \in X$ as $X$ is a subuniverse of $\alg{A}$. Therefore $t(b_0,\dotsc,b_{j-1}) \in X^+$ and we have shown that $X^+$ is closed under all term operations of $\alg{B}$, i.e. $X^+ \leq \alg{B}$.

Suppose $X$ absorbs $\alg{A}$ with respect to a term $t$ of arity $j$. Let $0\leq k < j$ be arbitrary and let $b_0,\dotsc,b_j \in B$ be elements such that $b_i \in X^+$ for all $i \neq k$. Then, for every $i, i \neq k$, we can find $a_i \in X$ such that $(a_i,b_i) \in R$. Also, since the projection of $R$ to the second coordinate is $B$, we can find $a_k \in A$ such that $(a_k,b_k) \in R$. We again have $(t(a_0,\dotsc,a_{j-1}), t(b_0,\dotsc,b_{j-1})) \in R$ and $t(a_0,\dotsc,a_{j-1}) \in X$~(as $X$ absorbs $\alg{A}$ with respect to $t$). It follows that $t(b_0,\dotsc,b_{j-1}) \in X^+$ and that $X^+\absorb \alg{B}$ with respect to $t$.

The remaining two statements are proved in an identical way.
\qed

\noindent
The subalgebra of $\alg{A}$ generated by $B$ can be obtained by applying term operations of $\alg{A}$ to elements of $B$.
The following auxiliary lemma provides a single term for all subsets $B$.

\begin{lem} \label{lem:subalgebraterm}
Let $\alg{A}$ be a finite idempotent algebra. Then there exists an operation $s \in \Clo(\alg{A})$ such that for any $B \subseteq A$ and any $b \in \Sg{\alg{A}}{B}$ there exists $a_0,\dotsc,a_{\arity{s}-1} \in B$ such that $s(a_0,\dotsc,a_{\arity{s}-1}) = b$.
\end{lem}

\proof
From the definition of $\Sg{\alg{A}}{B}$ it follows that
for every $B \subseteq A$ and every $b \in \Sg{\alg{A}}{B}$ there exists an operation $s_{(B,b)} \in \Clo(\alg{A})$ of arity $n$ and elements $a_0,\dotsc,a_{n-1}\in B$ such that $s_{(B,b)}(a_0,\dotsc, a_{n-1}) = b$. This operation is idempotent, as $\alg{A}$ is. 

For any two idempotent operations $t_1, t_2$ on $\alg{A}$~(of arities $n_1,\, n_2$) and any $a_0,\dotsc, a_{n_1-1}$, $b_0,\dotsc, b_{n_2-1}\in A$ we have
\begin{equation*}
t_1 \comp t_2 (
\underbrace{a_0,  \dots, a_0}_{n_2}, 
\underbrace{a_1,  \dots, a_1}_{n_2}, 
\dots,
\underbrace{a_{n_1-1},  \dots, a_{n_1-1}}_{n_2})
\end{equation*}
equal to $t_1(a_0,\dotsc,a_{n_1-1})$ and 
\begin{equation*}
t_1 \comp t_2 (b_0, b_1, \dots, b_{n_2-1}, \dotsc,b_0, b_1, \dots, b_{n_2-1}) 
\end{equation*}
equal to $t_2(b_0,\dotsc,b_{n_2-1})$.
Therefore the term operation
$$
s = s_{(B_1,b_1)} \comp s_{(B_2,b_2)} \comp \dots \comp s_{(B_l, b_l)},
$$
where $(B_1, b_1), (B_2, b_2), \dots, (B_l, b_l)$ is a complete list of pairs such that $b_i \in \Sg{\alg{A}}{B_i}$, satisfies the conclusion of the lemma.
\qed

\noindent
The following proposition is the only place in this article, where we use a Taylor term. Although the proof is quite easy, we believe that this proposition is of an independent interest. 

\begin{prop} \label{lem:bigterm}\label{prop:bigterm}
Let $\alg{A}$ be a finite algebra in a Taylor variety and suppose that $\alg{A}$ has no proper absorbing subalgebra.
Then there exists an operation $v \in \Clo(\alg{A})$ such that 
for any $b,c \in A$ and any coordinate $i < \arity{v}$ there exist $a_0,\dotsc,a_{\arity{v}-1} \in A$ such that $a_i = b$ and $v(a_0,\dotsc,a_{\arity{v}-1}) = c$.
\end{prop}

\proof
For a term operation $t \in \Clo(\alg{A})$ of arity $k$, an element $b \in A$, and a coordinate $i < \arity{t}$ we set
\begin{equation*}
W(t, b, i) = \{ t(a_0,\dotsc,a_{k-1}) : a_i = b \text{ and } \ a_j \in A\ \forall j \}.
\end{equation*}
Our aim is to find a term $v$ such that $W(v,b,i) = A$ for any $b \in A$ and any coordinate $i$.
We will achieve this goal by gradually enlarging the sets $W(t,b,i)$. 

Let $n < |A|$ and assume we already have an operation $v^{(n)} \in \Clo(\alg{A})$ such that each $W(v^{(n)},b,i)$ contains a subuniverse of $\alg{A}$ with at least $n$ elements. From idempotency it follows that all the one-element subsets of $A$ are subuniverses of $\alg{A}$, thus any operation in $\Clo(\alg{A})$ can be taken as $v^{(1)}$. 

For an induction step we first find an operation $w^{(n+1)} \in \Clo(\alg{A})$ such that each $W(w^{(n+1)},b,i)$ has at least $(n+1)$-elements:

\begin{claim}
Let $t \in \Clo(\alg{A})$ be a Taylor term operation and put $w^{(n+1)} = t \comp v^{(n)}$. Then $|W(w^{(n+1)},b,i)| > n$ for all $b \in A$ and all 
coordinates $i < \arity{w^{(n+1)}}$.
\end{claim}
\proof
Let $j = i \mdiv \arity{t}$, $k = i \mmod \arity{t}$ and let $B \subseteq W(v^{(n)},b,k)$ be a subuniverse of $\alg{A}$ with $|B| \geq n$.

First we observe that $B \subseteq W(w^{(n+1)},b,i)$. Indeed, take an arbitrary element $c \in B$, and find a tuple $a_0,\dotsc,a_{\arity{v^{(n)}}-1} \in A$ such that $a_k = b$ and that $v^{(n)}(a_0,\dotsc,a_{\arity{v^{(n)}}-1}) = c$. The application of $t\comp v^{(n)}$ to a concatenation of $\arity{t}$-many copies of $(a_0,\dotsc,a_{\arity{v^{(n)}}-1})$ produces $t(c, c, \dots, c) = c$. Since on the $i$-th coordinate of this catenation we have $b$, we showed that $c \in W(w^{(n+1)},b,i)$. Therefore if $B = A$ the claim holds and we can assume $B \varsubsetneq A$. 

As $t$ is a Taylor operation, it satisfies an identity of the form 
\begin{equation*}
t(\Box_0, \Box_1, \dots, \Box_{m-1}) \approx t(\triangle_0, \triangle_1, \dots, \triangle_{m-1}), 
\end{equation*}
where all $\Box_l$'s and $\triangle_l$'s are substituted with either $x$ or $y$, but $\Box_j$ is $x$ while $\triangle_j$ is $y$.

Let $r(x,y) = t(\Box_0, \Box_1, \dots, \Box_{m-1})$. Clearly $r \in \Clo(\alg{A})$. Since $\alg{A}$ has no proper absorbing subuniverses, the subuniverse $B$ is not an absorbing subuniverse of $\alg{A}$ with respect to the operation $r$. Therefore there exist $c \in B$ and $d \in A$ such that either
$r(c,d) \not\in B$ or $r(d,c) \not\in B$. We will show that $r(c,d), r(d,c) \in W(w^{(n+1)},b,i)$.

For each $e \in \{r(c,d), r(d,c)\}$ we can find a tuple $f_0,\dotsc,f_{\arity{t}-1} \in \{c,d\}$ such that $f_j = c$ and that $t(f_0,\dotsc,f_{\arity{t}-1}) = e$.
To obtain this we put 
\begin{iteMize}{$\bullet$}
 \item $f_l = c$ if $\Box_l = x$, and $f_l = d$ if $\Box_l = y$ in the case that $e = r(c,d)$ and 
 \item $f_l = c$ if $\triangle_l = x$, and $f_l = d$ if $\triangle_l = y$ in the case that $e = r(d,c)$.
\end{iteMize}
Further, since $c \in B \subseteq W(v^{(n)},b,k)$, we can find elements $a_0,\dotsc, a_{\arity{v^{(n)}}-1} \in A$ such that $a_k = b$ and $v^{(n)}(a_0,\dotsc, a_{\arity{v^{(n)}}-1}) = c$. To construct the argument for $t\comp v^{(n)}$ we expand each element of the tuple $(f_0,\dotsc, f_{\arity{t}-1})$ into $\arity{v^{(n)}}$-many identical copies of itself except $f_j$ which is substituted by $(a_0,\dotsc,a_{\arity{v^{(n)}}-1})$. It is easy to verify that $t\comp v^{(n)}$ applied to such an argument produces $e$.

We have proved that $B \cup \{r(c,d),r(d,c)\} \subseteq W(w^{(n+1)},b,i)$. As $|B|\geq n$ and  $r(c,d) \not\in B$ or $r(d,c) \not\in B$, we are done
\qed

\noindent
Now we are ready to define an operation $v^{(n+1)}$ such that each $W(v^{(n+1)},b,i)$ contains a subuniverse with at least $(n+1)$ elements:

\begin{claim}
Let $s$ be the operation from Lemma \ref{lem:subalgebraterm} and let $v^{(n+1)} = s \comp w^{(n+1)}$. Then, for all $b \in A$ and all coordinates $i < \arity{v^{(n+1)}}$,  $W(v^{(n+1)},b,i)$ contains a  subuniverse with more than $n$ elements. 
\end{claim}

\proof
Let $j = i \mdiv \arity{t}$, $k = i \mmod \arity{t}$ and let $B = W(w^{(n+1)},b,k)$. 
We will show that $\Sg{\alg{A}}{B} \subseteq W(v^{(n+1)},b,i)$.

Choose an arbitrary $c \in \Sg{\alg{A}}{B}$. By Lemma \ref{lem:subalgebraterm}, there exist  $f_0,\dotsc,f_{\arity{s}-1} \in B$ such that $s(f_0,\dotsc,f_{\arity{s}-1}) = c$. As before we prepare the tuple of arguments for $s\comp w^{(n)}$ by expanding the tuple $(f_0,\dotsc,f_{\arity{s}-1})$. Each $f_i$ gets expanded into $\arity{w^{(n+1)}}$-many identical copies of itself, except $f_j$ which gets expanded into a tuple $(a_0,\dotsc, a_{\arity{w^{(n+1)}}-1})\in A$ with $a_k=b$ and such that $w^{(n+1)}(a_0,\dotsc,a_{\arity{w^{(n+1)}}-1}) = f_j$~(such a tuple exists as $f_j\in B$). It is clear that $s\comp w^{(n+1)}$ applied to such a tuple produces $c$ and the claim is proved.
\qed

\noindent
To finish the proof of Proposition \ref{prop:bigterm}, it is enough to set $v = v^{(|A|)}$.
\qed

\noindent
It is an easy corollary that for two (or any finite number of) algebras in a Taylor variety we can find a common term satisfying the conclusion of Proposition~\ref{prop:bigterm}.

\begin{cor}
Let $\alg{A}, \alg{B}$ be finite algebras in a Taylor variety without proper absorbing subalgebras.
Then there exists a term $v$ such that 
for any $b,c \in A$ (resp. $b,c \in B$) and any coordinate $j < \arity{v}$ there exist $a_0,\dotsc,a_{\arity{v}} \in A$ (resp. $a_0,\dotsc,a_{\arity{v}} \in B$) such that $a_j = b$ and $v(a_0,\dotsc, a_{\arity{v}}) = c$.
\end{cor}

\proof
If $v_1$ (resp. $v_2$) is the term obtained from Proposition~\ref{prop:bigterm} for the algebra $\alg{A}$ (resp. $\alg{B}$), then we can put $v = v_1 \comp v_2$.
\qed

\noindent
We are now ready to prove Theorem~\ref{thm:abs}. 
One direction of the proof is straightforward: if an idempotent
variety $\variety{V}$ is not a Taylor variety, then, by
Theorem~\ref{thm:taylor}, it contains a two-element algebra
whose every operation is a projection. 
Such an algebra has no absorbing
subuniverses and any three-element subset of its square is a linked
subdirect subalgebra which falsifies the second condition of Theorem~\ref{thm:abs}. Therefore it remains to prove the following. 
\begin{thm} \label{thm:absreal}
Let $\alg{A}, \alg{B}$ be finite algebras in a Taylor variety and let $R$ be a proper, subdirect and linked subalgebra of $\alg{A} \times \alg{B}$. Then $\alg{A}$ or $\alg{B}$ has a proper absorbing subalgebra.
\end{thm}
\noindent

\proof
For contradiction, assume that $R, \alg{A}, \alg{B}$ form a counterexample to the theorem. Thus neither $\alg{A}$ nor $\alg{B}$ has a proper absorbing subalgebra and 
$R \leq_S \alg{A} \times \alg{B}$ is a linked, proper subset of $A \times B$. 

First we find another counterexample satisfying $R^{-1} \circ R = A \times A$. 
As $R$ is  linked, there exists a natural number $k$ such that $(R^{-1} \circ R)^{ \circ k} = A^2$. Take the smallest such $k$. 
If $k=1$, then $R^{-1} \circ R = A \times A$ and we need not to do anything. Otherwise we replace $\alg{B}$ by $\alg{A}$ and $R$ by
$(R^{-1} \circ R)^{ \circ (k-1)}$. Our new choice of $R, \alg{A}, \alg{B}$ is clearly a counterexample to the theorem satisfying $R^{-1} \circ R = A \times A$.

From now on we assume that our counterexample satisfies $R^{-1} \circ R = A \times A$. In other words, for any $a, c \in A$, there exists $b \in B$ such that $(a,b), (c,b) \in R$.

For a $X \subseteq A$  we set
\begin{eqnarray*}
N(X) &=& \{b \in B: \forall \ a \in X \ \ (a,b) \in R\} = \bigcap_{a \in X} \{a\}^+
\end{eqnarray*}

\begin{claim}\label{claim:NSG}
$N(X) = N(\Sg{\alg{A}}{X})$.
\end{claim}

\proof
If $t$ is a $k$-ary term, $a_0,\dotsc,a_{k-1}$ are elements of $X$ and $b \in N(X)$, then $(a_i, b) \in R$ for any $i=0,1, \dots, k-1$. Therefore
$(t(a_0,\dotsc,a_{k-1}),b) \in R$. This shows that $b \in \{t(a_0,\dotsc,a_{k-1})\}^+$.
\qed

\begin{claim}
 $N(A)\neq \emptyset$.
\end{claim}
\proof
We call a subset $X \subseteq A$ \emph{good}, if $(N(X))^{-} = A$. Since $R^{-1} \circ R = A \times A$, every one-element subset of $A$ is good. We prove the claim by showing that $A$ is good.

Let $X$ be  a maximal, with respect to inclusion, good subset of $A$. 
We know that $\emptyset \neq X$, since each one-element subset is good, and also $X \neq A$, otherwise the claim is proved. 
As $N(X) = N(\Sg{\alg{A}}{X})$ due to the Claim~\ref{claim:NSG}, $X$ is a subuniverse of $\alg{A}$. Let $v \in \Clo(\alg{A})$ be the operation from
Proposition \ref{lem:bigterm}. Due to our assumption that $\alg{A}$ has no proper absorbing subuniverses, $X$ is not an absorbing subuniverse of $\alg{A}$ with respect to the operation $v$. It follows that there exists a coordinate $j < \arity{v}$ and elements $a_0,\dotsc,a_{\arity{v}-1} \in A$ such that
$a_i \in X$ for all $i \neq j$, and $b:=v(a_0,\dotsc, a_{\arity{v}-1}) \not\in X$. 

We will prove that the set $X \cup \{b\}$ is good, which will contradict the maximality of $X$. Let $c \in A$ be arbitrary. From  Proposition \ref{lem:bigterm}
we obtain $d_0,\dotsc,d_{\arity{v}-1} \in A$ such that $d_j = a_j$ and $v(d_0,\dotsc,d_{\arity{v}-1}) = c$. Since $(N(X))^{-}=A$, we can
find $e_0,\dotsc,e_{\arity{v}-1} \in N(X)$ such that  $(d_i,e_i) \in R$ for all $i$. Put $f = v(e_0,\dotsc,e_{\arity{v}-1})$. As $R$ is a subuniverse of $\alg{A} \times \alg{B}$ and $(d_i,e_i) \in R$ for all $i$, it follows that $(v(d_0,\dotsc,d_{\arity{v}-1}),v(e_0,\dotsc,e_{\arity{v}-1})) = (c,f) \in R$. The set $N(X)$ is a subuniverse of $\alg{B}$ thus we have $f \in N(X)$. For all $i \neq j$, we have $a_j \in X$ and $e_j \in N(X)$, 
hence $(a_j,e_j) \in R$. But also $(a_i=d_i,e_i) \in R$ and, again, $R$ is a subuniverse of $\alg{A} \times \alg{B}$, therefore $(v(a_0,\dotsc,a_{\arity{v}-1}),v(e_0,\dotsc,e_{\arity{v}-1})) = (b,f) \in R$. We have proved that, for any $c \in A$, there exists $f \in N(X) \cap \{b\}^+ = N(X \cup \{b\})$ such that $(c,f) \in R$. Therefore $X \cup \{b\}$ is good, a contradiction. This contradiction shows that $N(A)$ is nonempty.
\qed 
Since $R$ is a proper subset of $A \times B$, $N(A)$ is a proper subset of $B$. This set is an intersection of subuniverses of $\alg{B}$, thus $N(A)$ a subuniverse of $\alg{B}$.
Since $N(A)$ is not an absorbing subuniverse of $\alg{B}$ with respect to $v$, there exists a coordinate $j < \arity{v}$ and a tuple
$b_0,\dotsc,b_{\arity{v}-1} \in B$ such that $b_i \in N(A)$ for all $i \neq j$, and $c:=v(b_0,\dotsc,b_{\arity{v}-1}) \not\in N(A)$. 

We will prove that $(d,c) \in R$ for all $d \in A$, which will contradict the definition of $N(A)$. Let $a \in A$ be any element of $A$ such that $(a,b_j) \in R$ (we use subdirectness of $R$ here) and let $a_i \in A$ be obtained from Proposition \ref{lem:bigterm} in  such a way that $a_j = a$ and $v(a_0,\dotsc,a_{\arity{v}})=d$.
For all $i \neq j$, we have $(a_i,b_i) \in R$ as $b_i \in N(A)$, and also $(a_j=a,b_j) \in R$. Thus $(v(a_0,\dotsc,a_{\arity{v}-1}),v(b_0,\dotsc,b_{\arity{v}-1})) = (d,c) \in R$.
\qed

\subsection{Minimal absorbing subalgebras}
We present a number of properties of absorbing subuniverses required in the proof of Theorem~\ref{thm:cyclic}.
Most of them are corollaries of the Absorption Theorem and they give us some information about minimal absorbing subalgebras:
\begin{defi}
If $\alg{B} \absorb \alg{A}$ and no proper subalgebra of $\alg{B}$ absorbs $\alg{A}$, we call $\alg{B}$ a \emph{minimal absorbing subalgebra} of $\alg{A}$~(and write $\alg{B} \minabs \alg{A}$).
\end{defi}
\noindent Alternatively, we can say that $\alg{B}$ is a minimal absorbing subalgebra of $\alg{A}$, if $\alg{B} \absorb \alg{A}$ and $\alg{B}$ has no proper absorbing subalgebras. Equivalence of these definitions follows from transitivity of $\absorb$ (proved in Proposition \ref{prop:trans}). 
Observe also that two minimal absorbing subuniverses of $\alg{A}$ are either disjoint or coincide, but the union of all minimal absorbing subuniverses need not be the whole set $A$.

\begin{prop} \label{prop:abscor}
Let $\variety{V}$ be a Taylor variety, let $\alg{A}$ and $\alg{B}$ be finite algebras in $\variety{V}$ and let $\alg{R} \leq_S \alg{A} \times \alg{B}$.
\begin{enumerate}[\em(i)]
\item If $R$ is linked and $\alg{E} \absorb \alg{R}$, then $E$ is linked.
\item If $\alg{C} \minabs \alg{A}$, $\alg{D} \minabs \alg{B}$, and $(C \times D) \cap R \neq \emptyset$, then $(\alg{C} \times \alg{D}) \cap R \leq_S \alg{C} \times \alg{D}$.
\item If $R$ is linked, $\alg{C} \minabs \alg{A}$, $\alg{D} \minabs \alg{B}$, and $(C \times D) \cap R \neq \emptyset$, then $\alg{C} \times \alg{D} \minabs \alg{R}$.
\item If $R$ is linked, and $\alg{C} \minabs \alg{A}$, then there exists $\alg{D} \minabs \alg{B}$ such that $C \times D \subseteq R$.
\item 
If $R$ is linked, $\alg{C} \minabs \alg{A}$~or $\alg{C} \minabs \alg{B}$, $\alg{D} \minabs \alg{A}$~or $\alg{D} \minabs \alg{B}$, $c \in C$, and $d \in D$, then $c$ and $d$ can be linked via $c_0,\dotsc,c_{j}$ where each $c_{i}$ is a member of some minimal absorbing subalgebra of $\alg{A}$ or $\alg{B}$.
\end{enumerate}
\end{prop}

\noindent
To avoid ambiguity in the statement of item (v), assume that the algebras $\alg{A},\alg{B}$ are disjoint. When we apply the corollary this need not be the case, but the assumptions~(and therefore conclusions) of the corollary will be satisfied when we substitute the algebras $\alg{A},\alg{B}$ with their isomorphic, disjoint copies.

\proof \hfill

\begin{enumerate}[(i)]
\item
Suppose that $\alg{E}$ absorbs $\alg{R}$ with respect to an operation $t$. Let $(a,b), (a',b')$ be arbitrary elements of $E$. As $R$ is linked, there exist $c_0, c_1, \dots, c_{2n} \in A \cup B$ such that $c_0=a$, $c_{2n}=a'$, $(c_{2i},c_{2i+1}) \in R$ and $(c_{2i+2},c_{2i+1}) \in R$ for all $i=0,1, \dots, n-1$. The pair 
\begin{equation*}
t((c_{2i},c_{2i+1}),(a,b), (a,b), \dots, (a,b)), 
\end{equation*}
which is, by definition of the product of two algebras, equal to
\begin{equation*}
(t(c_{2i},a,a, \dots, a), t(c_{2i+1},b,b, \dots, b))
\end{equation*}
is in $E$ for all $i$, since $E$ absorbs $R$ with respect to $t$. Similarly, 
\begin{equation*}
(t(c_{2i+2}, a, a, \dots, a),t(c_{2i+1}, b, b, \dots, b)) \in E.
\end{equation*}
Therefore the elements $a = t(a, a, \dots, a)$ and $t(a',a,a, \dots a)$ are linked in $E$ via
$t(c_0, a, \dots, a)$, $t(c_1, b, \dots, b)$, \dots, $t(c_{2n}, a, \dots, a)$.

Using the same reasoning,
the pairs 
\begin{equation*}
(t(a',c_{2i}, a, \dots, a), t(b',c_{2i+1}, b, \dots, b))
\end{equation*}
and 
\begin{equation*}
(t(a',c_{2i+2}, a, \dots, a), t(b',c_{2i+1}, b, \dots, b)) 
\end{equation*}
are in $E$ and it follows that $t(a',a,a, \dots, a)$ and $t(a',a',a,a, \dots, a)$ are linked in $E$.  By continuing similarly we get that $a = t(a,a, \dots, a)$ and $a' = t(a',a', \dots, a')$ are linked in $E$ as required.

\item
By Lemma~\ref{lem:neig} $\alg{D}^-\absorb \alg{A}$, therefore $\emptyset \neq (\alg{D}^-\cap\alg{C})\absorb \alg{A}$ (by Proposition \ref{prop:trans})  and, as $\alg{C}\minabs\alg{A}$, we get $\alg{D}^-\supseteq \alg{C}$. A symmetric reasoning shows that $\alg{C}^+\supseteq \alg{D}$ and the item is proved.

\item
 Let $E = (C \times D) \cap R$ and let $\alg{E}$ be the subalgebra of $\alg{A} \times \alg{B}$ with universe $E$. 
 From (ii) it follows that $E \leq_S \alg{C} \times \alg{D}$.
Clearly $E \absorb \alg{R}$, therefore $E$ is linked by (i). Theorem \ref{thm:absreal} together with the minimality of $\alg{C}$ and $\alg{D}$ now gives $E = C \times D$.

Let $\emptyset \neq F \absorb \alg{E}$. The projection of $F$ to the first (resp. the second) coordinate is clearly an absorbing subuniverse of 
$\alg{C}$ (resp. $\alg{D}$). Therefore $F \leq_S \alg{C} \times \alg{D}$. Using (i) and Theorem \ref{thm:absreal} as above we conclude that $F = C \times D$.

\item
Let $D' = C^{+}$. According to Lemma \ref{lem:neig}, $D'$ is an absorbing subuniverse of $\alg{B}$. Let $\alg{D}'$ be the subalgebra of $\alg{A}$ with universe $D'$ and let $\alg{D}$ be a minimal absorbing subalgebra of $\alg{D}'$. The claim now follows from (iii).

\item
We prove this fact by induction on the length of the path connecting $c$ and $d$.  If the length is $2$, then we have $c,d \in A$ (thus $\{c\}^{+}\cap\{d\}^+ \neq\emptyset$), or $c,d \in B$ (thus $\{c\}^-\cap\{d\}^- \neq\emptyset$). Without loss of generality we assume the first case and, conclude using Lemma~\ref{lem:neig} and Proposition~\ref{prop:trans}, that $\emptyset\neq (\alg{C}^+\cap\alg{D}^+)\absorb \alg{B}$. Let $E$ be any subuniverse such that $E\minabs (\alg{C}^+\cap\alg{D}^+)$. Then, as $(C\times E)\cap R\neq \emptyset$ and $(D\times E)\cap R \neq \emptyset$, by (iii), we obtain $C\times E\subseteq R$ and $D\times E\subseteq R$ and the first case is proved.

For the induction step, we assume, without loss of generality, that $\alg{C}\minabs \alg{A}$ and define $C_0 = C, C_1= C_0^+, C_2 = C_1^-, C_3=C_2^+,\dotsc$ with $d\in C_n$. Suppose, for simplicity of the presentation, that $d$ appears on the right side (i.e. $d \in B$) and consider $(C_{n-1}\cap\alg{D}^-) \absorb \alg{A}$. Let $\alg{E}\minabs (C_{n-1}\cap\alg{D}^-)$. By (iii) we have $E\times D\subseteq R$ and, by inductive assumption we have an element of $E$, say $e$, linked inside minimal absorbing subuniverses to some element of $\alg{C}$ say $c'$. Therefore $d$ is linked~(through $e$) inside minimal sets to some $c'\in\alg{C}$. By (iv) we link, inside minimal absorbing subuniverses, $c'$ to $c$ and the item is proved.
\end{enumerate}
\qed

\section{New proof of the Smooth Theorem}\label{sect:smoothproof}

\noindent
The Smooth Theorem classifies the computational complexity of CSPs generated by smooth digraphs~(digraphs, where every vertex has at least one incoming and at least one outgoing edge). This classification was conjectured by Bang-Jensen and Hell~\cite{BJH} and confirmed by the authors in~\cite{smoothproc,smoothpaper}. The proof presented in those papers heavily relied on the results of McKenzie and Maroti~\cite{MM} which characterized the locally finite Taylor varieties in terms of weak near-unanimity operations. 
We present an alternative proof which depends only on Theorem~\ref{thm:abs}. The Smooth Theorem states:
\begin{thm}\label{thm:smooth}
Let $\graph{H}$ be a smooth digraph. If each component of the core of $\graph{H}$ is a circle, then $\CSP(\graph{H})$ is polynomially
decidable. Otherwise $\CSP(\graph{H})$ is NP-complete.
\end{thm}

\subsection{Basic digraph notions}
\noindent A \emph{digraph} is a pair $\graph{G} = (V,E)$, where $V$ is a finite set of vertices and $E \subseteq V \times V$ is a set of edges.
If the digraph is fixed we write $a \rightarrow b$ instead of $(a,b) \in E$. 
The induced subgraph of $\graph{G}$ with vertex set $W \subseteq V$ is denoted by $\indg{\graph{G}}{W}$, that is, 
$\indg{\graph{G}}{W} = (W, E \cap (W \times W))$.
A \emph{loop} is an edge of the form $(a,a)$.
$\graph{G}$ is said to be \emph{smooth} if every vertex has an incoming and an outgoing edge, in other words, $\graph{G}$ is smooth, if $E$ is a subdirect product of $V$ and $V$. The \emph{smooth part} of $\graph{G}$ is the largest subset $W$ of $V$ such that $\indg{\graph{G}}{W}$ is  smooth~(it can be empty).

An \emph{oriented path} is a digraph $\relstr{P}$ with vertex set $P = \{p_0, \dots, p_k\}$ and edge set consisting of $k$ edges --- for all $i < k$ either $(p_i, p_{i+1})$, or $(p_{i+1},p_i)$ is an edge of $\graph{P}$. An \emph{initial segment} of such a path is any path induced by $\relstr{P}$ on vertices $\{p_0,\dotsc,p_i\}$ for some $i<k$. We denote the oriented path consisting of $k$ edges pointing forward by $\cdot\xrightarrow{k}\cdot$ and, similarly the oriented path consisting of $k$ edges pointing backwards by $\cdot\xleftarrow{k}\cdot$. The concatenation of paths is performed in the natural way. A \emph{$(k,n)$-fence}~(denoted by $\fence{k}{n}$) is the oriented path consisting of $2kn$ edges, $k$ forward edges followed by $k$ backward edges, $n$ times i.e.:
\begin{equation*}
 \underbrace{\cdot\xrightarrow{k}\cdot\xleftarrow{k}\cdots\cdot\xrightarrow{k}\cdot\xleftarrow{k}\cdot}_{n}
\end{equation*}

The \emph{algebraic length} of an oriented path is the number of forward edges minus the number of backward edges~(and thus all the fences have algebraic length zero).  Let $\graph{G}$ be a digraph, let $\graph{P}$ be an oriented path with vertex set $P = \{p_0, \dots, p_k\}$, and let $a,b$ be vertices of $\graph{G}$. We say that
$a$ is \emph{connected to} $b$ via $\relstr{P}$,
if there exists a homomorphism $f: \graph{P} \rightarrow \graph{G}$ such that $f(p_0) = a$ and $f(p_k) = b$. We sometimes write $a\xrightarrow{k} b$ when $a$ is connected to $b$ via $\cdot\xrightarrow{k}\cdot$. If $a\xrightarrow{k}a$~(for some $k$) then $a$ is \emph{in a cycle} and any image of the path $\cdot\xrightarrow{k}\cdot$ with the same initial and final vertex is \emph{a cycle}. \emph{A circle} is a cycle which has no repeating vertices and no chords.

The relation ``$a$ is connected to $b$~(via some path)'' is an equivalence, its blocks (or sometimes the corresponding induced subdigraphs) are called the \emph{weak components} of $\graph{G}$.  The vertices $a$ and $b$ are in the same \emph{strong component} if $a\xrightarrow{k}b\xrightarrow{k'}a$ for some $k,k'$. For a subset $B$ of $A$ and an oriented path $\relstr{P}$ we set
\begin{equation*}
B^{\relstr{P}} = \{ c : \exists b \in B \ \mbox{ $b$ is connected to $c$ via $\relstr{P}$ } \}.
\end{equation*}
Note that $B^{\cdot\xrightarrow{k}\cdot}$ is formally equal to $B^{+E^{\circ k}}$ but we prefer the first notation.

Finally, $\graph{G}$ has \emph{algebraic length $k$}, if there exists a vertex $a$ of $\graph{G}$ such that $a$ is connected to $a$ via a path of algebraic length $k$ and $k$ is the minimal positive number with this property. The following proposition summarizes easy results concerning reachability via paths:

\begin{prop}\label{prop:basicsmooth}
Let $\graph{G}$ be a smooth digraph, then:
\begin{iteMize}{$\bullet$}
 \item for any vertices $a,b$ in $\graph{G}$ if $a$ is connected to $b$ via $\cdot\xrightarrow{k}\cdot$ then $a$ is connected to $b$ via every path of algebraic length $k$;
 \item for any vertex $a$ and any path $\relstr{P}$ there exists a vertex $b$ and a path $\relstr{Q}$ which is an initial segment of some fence such that $\{a\}^{\graph{P}}\subseteq\{b\}^{\graph{Q}}$;
 \item if $H\subseteq G$ is such that $H^{\cdot\rightarrow\cdot}\supseteq H$ or $H^{\cdot\leftarrow\cdot}\supseteq H$ then the digraph $\indg{\graph{G}}{H}$ contains a cycle~(i.e. \emph{the smooth part} of $\indg{\graph{G}}{H}$ is non-empty)
\end{iteMize}
\end{prop}
\proof
The first item of the proposition follows directly form the definition of a smooth digraph.

 We prove the second item by induction on the length of $\graph{P}$. If the length is zero there is nothing to prove. Therefore we take an arbitrary path $\relstr{P}$ of length $n$ and arbitrary $a\in A$. The proof splits into two cases depending on the direction of the last edge in $\graph{P}$. We consider the case when the last edge of $\graph{P}$ points forward first and set $\graph{P'}$ to be $\graph{P}$ take away the last edge. The inductive assumption for $a$ and $\relstr{P'}$ provides a vertex $b$ and a path $\relstr{Q'}$~(an initial fragment of a fence $\fence{k}{l}$). If the algebraic length of $\graph{Q'}$ is strictly smaller than $k$, we put $\graph{Q'''}$ to be a path such that the concatenation of $\graph{Q'}$ and $\graph{Q'''}$ is an initial fragment of the fence $\fence{k}{l+1}$ and such that the algebraic length of $\graph{Q'''}$ is one; then the concatenation of $\graph{Q'}$ and $\graph{Q'''}$ proves the second item of the proposition~(as, by the first item of the proposition, every element reachable from $\{b\}^{\graph{Q'}}$ by $\cdot\rightarrow\cdot$ is also reachable by $\graph{Q'''}$). If the algebraic length of $\graph{Q'}$ equals $k$ we consider a path $\graph{Q''}$ obtained from $\graph{Q'}$ by substituting each subpath of the shape $\cdot\rightarrow\cdot\leftarrow\cdot$ with $\cdot\xrightarrow{2}\cdot\xleftarrow{2}\cdot$. The path $\graph{Q''}$ is an initial fragment of $\fence{k+1}{l}$ and we have $\{b\}^{\graph{Q'}}\subseteq\{b\}^{\graph{Q''}}$~(as the digraph is smooth). Now we can find $\graph{Q'''}$ as in the previous case.

If the last edge of $\graph{P}$ points backwards, we proceed with dual reasoning. If the algebraic length of $\graph{Q'}$ is greater than zero we obtain $\graph{Q'''}$ of algebraic length $-1$ as before and the proposition is proved. If the algebraic length of $\graph{Q'}$ is zero we substitute $b$ with any vertex $b'$ such that $b'\rightarrow b$ and alter $\graph{Q'}$ by substituting each $\cdot\leftarrow\cdot\rightarrow\cdot$ with $\cdot\xleftarrow{2}\cdot\xrightarrow{2}\cdot$. The new path is an initial fragment of $\fence{k+1}{l}$ and we can proceed as in previous case.

For the third item of the proposition. Without loss of generality we can assume the first possibility and choose an arbitrary $b_0\in H$. As $H\subseteq H^{\cdot\rightarrow\cdot}$ there is an element $b_1\in H$ such that $b_1\rightarrow b_0$. Repeating the same reasoning for $b_1, b_2,\dotsc$ we obtain a sequence of vertices in $H$ such that $b_{i+1}\rightarrow b_i$. As $H$ is finite, we obtain a cycle in $H$ and the last item of the proposition is proved.
\qed

\noindent
The following lemma shows that the smooth part of an induced subdigraph of a smooth digraph shares some algebraic properties with the induced subdigraph.

\begin{lem}\label{lem:bothways}
Let $\alg{A}$ be a finite algebra and let $\relstr{G} = (A,E)$ be a smooth digraph such that $E$ is a subuniverse of $\alg{A}^2$. If $B$ is a subuniverse of $\alg{A}$~(an absorbing subuniverse of $\alg{A}$) then the smooth part of $\indg{\graph{G}}{B}$ forms a subuniverse of $\alg{A}$~(an absorbing subuniverse of $\alg{A}$ respectively).
\end{lem}
\proof
 Note that if the smooth part of $\indg{\graph{G}}{B}$ is empty then the lemma holds. Assume it is non-empty and let $\alg{A}$, $\graph{G}$, $B$ be as in the statement of the lemma. We put $B_1\subseteq B$ to be the set of all the vertices in $B$ with at least one outgoing and at least one incoming edge in $\indg{\graph{G}}{B}$~(i.e. an outgoing edge and an incoming edge to elements of $B$). As $B_1 = B\cap B^{+E}\cap B^{-E}$ Lemma~\ref{lem:neig} implies that $B_1$ is a subuniverse~(absorbing subuniverse resp.) of $\alg{A}$. We put $B_2= B_1\cap B_1^{+E}\cap B_1^{-E}$ and continue the reasoning. Since $\alg{A}$ is finite we obtain some $k$ such that $B_k=B_{k+1}$. Since $\indg{\graph{G}}{B_k}$  has no sources and no sinks the lemma is proved.
\qed

\subsection{Reduction of the problem}
\noindent 
The first part of Theorem~\ref{thm:smooth} is easy: if a digraph $\graph{H}$ has a core which is a disjoint union of circles then $\CSP(\graph{H})$ is solvable in polynomial time (see \cite{BJH}). On the other hand, using Theorem~\ref{thm:bjk} and the fact that
CSPs of a relational structure and its core are the same, it suffices to prove that:

\begin{thm}\label{thm:smoothintermediate}
If a smooth digraph admits a Taylor polymorphism then it retracts onto the disjoint union of circles. 
\end{thm}

\noindent 
Finally, Theorem~\ref{thm:smoothintermediate} reduces to the theorem below. An elementary proof of this reduction can be found in~\cite{smoothproc,smoothpaper}.

\begin{thm}\label{thm:oldsmooth}
If a smooth digraph has algebraic length one and admits a Taylor polymorphism then it contains a loop.
\end{thm}

\noindent In fact, in the remainder of this section, we prove a stronger version of Theorem~\ref{thm:oldsmooth}:
\begin{thm}\label{thm:realsmooth}
Let $\alg{A}$ be a finite algebra in a Taylor variety and let $\relstr{G} = (A,E)$ be a smooth digraph of algebraic length one such that $E$ is a subuniverse of $\alg{A}^2$. Then $\relstr{G}$ contains a loop. Moreover, if  there exists an absorbing subuniverse $I$ of $\alg{A}$ which is contained in a weak component of $\relstr{G}$ of algebraic length $1$, then the loop can be found in some  $J$ such that $J \minabs \alg{A}$.
\end{thm}

\subsection{The proof}
Our proof of Theorem~\ref{thm:realsmooth} proceeds by induction on the size of the vertex set of $\graph{G}=(A,E)$. If $|A|=1$ there is nothing to prove~(as the only smooth digraph on such a set contains a loop); for the induction step we assume that Theorem~\ref{thm:realsmooth} holds for all smaller digraphs. 
\begin{claim}\label{claim:abs1}
 Let $H$ be a weak component of $\graph{G}$ of algebraic length one, then there exists $a\in H$ and a path $\relstr{P}$ such that $\{a\}^{\relstr{P}}$ contains a cycle.
\end{claim}
\proof
 We choose $a\in H$ to be the element of the component $H$ such that there is a path $\relstr{Q}$ of algebraic length one connecting $a$ to $a$. We define the sequence of sets $B_0=\{a\}$ and $B_i = B_{i-1}^{\relstr{Q}}$ recursively. As $a$ is connected to $a$ via $\relstr{Q}$ we have $B_0\subseteq B_1$ and therefore $B_i\subseteq B_{i+1}$ for any $i$~(as by definition $B_{i-1}\subseteq B_i$ implies that $B_{i-1}^{\relstr{Q}} \subseteq B_i^{\relstr{Q}}$ i.e. $B_i\subseteq B_{i+1}$). As $\relstr{Q}$ is of algebraic length one we can use Proposition~\ref{prop:basicsmooth} to infer that $\{a\}^{\cdot\rightarrow\cdot}\subseteq B_1$ and further that $\{a\}^{\cdot\xrightarrow{k}\cdot}\subseteq B_k$ for any $k$. These facts together imply that
\begin{equation*}
 \bigcup_{i=0}^{k} \{a\}^{\cdot\xrightarrow{i}\cdot} \subseteq B_k
\end{equation*}
and, as the digraph is finite, we can find a cycle in one of the $B_k$'s. Take $\relstr{P}$ to be the $\relstr{Q}$ concatenated with itself sufficiently many times to witness the claim. 
\qed
\begin{claim}\label{claim:abs2}
 Let $H$ be a weak component of $\graph{G}$ of algebraic length one, then there exists $a\in H$ and a fence $\Fence$ such that $\{a\}^{\Fence}=H$.
\end{claim}
\proof
Let us choose $a\in H$ and $\relstr{P'}$ as provided by Claim~\ref{claim:abs1}. Set $B$ to be the set of elements of $\{a\}^{\relstr{P'}}$ which belong to some cycle fully contained in $\{a\}^{\relstr{P'}}$. Proposition~\ref{prop:basicsmooth} implies that $B^{\fence{|A|}{1}}$ contains all elements reachable by $\cdot\xrightarrow{i}\cdot$ or $\cdot\xleftarrow{i}\cdot$~(for any $i$), from any element of $B$. Indeed if such a $c$ is reachable from $b\in B$ by $\cdot\xleftarrow{i}\cdot$ then it is reachable by $\cdot\xleftarrow{|A|}\cdot$ from some $b'\in B$ and further by $\fence{|A|}{1}$ from some $b''\in B$. In the other case $b\xrightarrow{i}c$ for some $b\in B$. There obviously exists $d$ such that $d\xleftarrow{|A|} c$ and since $b\xrightarrow{i} c \xrightarrow{|A|} d$ we have some $j\leq |A|$ and $b\xrightarrow{j} d$. Thus there exists $b'\in B$ with $b'\xrightarrow{|A|} d$ and $c$ is reachable by $\fence{|A|}{1}$ from $b'$.

For every element $c$ in $H$ we can find $b_0, b_1, \dots, b_{|A|} = c$ such that 
each $b_j$, $j \neq |A|$, is in a cycle $B_j$ where $B_0 \subseteq B$, and 
 $b_0 \xrightarrow{i_0} b_1 \xleftarrow{i_1} b_2 \xrightarrow{i_2} b_3 \xleftarrow{} \dots  b_{|A|}$ for some $i_0, i_1, \dots, i_{|A|-1}$.
The reasoning above shows that $B_j$ is contained in $B_{j-1}^{\fence{|A|}{1}}$ (for all $1 \leq j < |A|$) and $b_{|A|}$ belongs to $B_{|A|-1}^{\fence{|A|}{1}}$, therefore  
$B^{\fence{|A|}{|A|}} = H$.

Thus, for an appropriate path $\relstr{P}$ we have $a$ connected to every element of $H$ by $\relstr{P}$. The second item of Proposition~\ref{prop:basicsmooth} provides $b$ and an initial segment $\graph{Q}$ of a fence $\Fence$ such that $b$ is connected to every element from $H$ by $\graph{Q}$. 
Let $\relstr{S}$ denote the remaining part of the fence $\Fence$. Then $\{b\}^{\Fence} = (\{b\}^{\graph{Q}})^{\relstr{S}} = H^{\relstr{S}} = H$ 
and the claim is proved. 
\qed

\noindent
The remaining part of the proof splits into two cases: in the first case the algebra $\alg{A}$ has an absorbing subuniverse in a weak component of algebraic length one and in the second it doesn't. Let us focus on the first case and define $I\absorb \alg{A}$ contained in a weak component~(denoted by $H$) of algebraic length one  of $\graph{G}$.
\begin{claim}\label{claim:1}
 There is a fence $\Fence$ such that $I^{\Fence} = H$.
\end{claim}
\proof
 Let $a$ and $\Fencepr$ be provided by Claim~\ref{claim:abs2}. We put $\Fence$ to be a concatenation of $\Fencepr$ with itself. Since $a\in I^{\Fencepr}$, then $I^{\Fence} = H$.
\qed

\noindent Let $\graph{P}$ be the longest initial segment of $\Fence$~(provided by Claim~\ref{claim:1}) such that $I^{\graph{P}}\neq H$. Put $S=I^{\graph{P}}$.  By multiple application of Lemma~\ref{lem:neig} we infer that $S$ is a subuniverse of $\alg{A}$ and that $S\absorb\alg{A}$. 
%
%
The definition of $S$ implies that $S^{\cdot\rightarrow\cdot}=H\supseteq S$ or $S^{\cdot\leftarrow\cdot}=H\supseteq S$, and therefore, by Proposition~\ref{prop:basicsmooth}, $S$ contains a cycle. Thus the smooth part of $\indg{\graph{G}}{S}$, denoted by $S'$, is non-empty and, by Lemma~\ref{lem:bothways}, it absorbs $\alg{A}$. If the digraph $\indg{\graph{G}}{S'}$ has algebraic length one and is weakly connected, then we use the inductive assumption: 
\begin{iteMize}{$\bullet$}
\item either $\indg{\graph{G}}{S'}$ has no absorbing subuniverses in a weak component of algebraic length one; in such a case, as it is weakly connected, it has no absorbing subuniverses at all ---  therefore $S'\minabs \alg{A}$ and the inductive assumption provides a loop in $S'$, or
\item $\indg{\graph{G}}{S'}$ has an absorbing subuniverse; then it has a loop in $J\minabs S'$ and, as $J\minabs\alg{A}$, the theorem is proved. 
\end{iteMize}
Therefore to conclude the first case of the theorem it remains to prove

\begin{claim}
$\indg{\graph{G}}{S'}$ is a weakly connected digraph of algebraic length $1$.
\end{claim}

\proof
Assume that $S'$ absorbs $\alg{A}$ with respect to $t$ of arity $k$ and
let $m,n$ be natural numbers such that every two vertices of $H$ are connected via the $(m,n)$-fence~(implied by Claim~\ref{claim:abs2}) denoted by $\relstr{F}$. We will show that any two vertices $a,b \in S'$ are connected via the $(m,nk)$-fence in the digraph $\indg{\graph{G}}{S'}$.

As the digraph $\indg{\graph{G}}{S'}$ is smooth, $a$ is connected to $a$ via $\relstr{F}$ and $b$ is connected to $b$ via $\relstr{F}$~(by the first item of Proposition~\ref{prop:basicsmooth}). Let $f: \relstr{F} \rightarrow S'$ and $g: \relstr{F} \rightarrow S'$ be the corresponding digraph homomorphisms. Moreover, $a$ is connected to $b$ via $\relstr{F}$ in the digraph $\relstr{G}$ and we take the corresponding homomorphism $h: \relstr{F} \rightarrow \relstr{G}$.
For every $i = 0,1, \dots, k-1$ we consider the following matrix with $k$ rows and $2nm+1$ columns: 
To the first $(k-i-1)$ rows we write $f$-images of the vertices of $\relstr{F}$, to the $(k-i)$th row we write $h$-images, and to the last $i$ rows we write $g$-images. We apply the term operation $t$ to columns of this matrix. Since $E \leq \alg{A}^2$ we obtain a homomorphism from $\relstr{F}$ to $\relstr{G}$ which realizes a connection from 
$$t(\underbrace{a, a, \dots, a}_{(k-i)}, \underbrace{b, b, \dots, b}_{i})$$ 
to
$$t(\underbrace{a, a, \dots, a}_{(k-i-1)}, \underbrace{b, b, \dots, b}_{(i+1)}).$$
Moreover, since all but one member of each column are elements of $S'$ and $S' \absorb \alg{A}$, we actually get a
homomorphism $\relstr{F} \rightarrow S'$.
By joining these homomorphisms for $i=0,1, \dots, k-1$ we obtain that $a = t(a,a, \dots, a)$ is connected to $b = t(b,b, \dots, b)$ via the $(m,nk)$-fence in $S'$.

As $S'\subseteq H$ all the elements of $S'$ are connected in $H$, and, using the paragraph above, also in $S'$. Moreover we can take two elements $a,b\in S'$ such that $a\rightarrow b$. As $a$ is connected to $b$ via a $(m,nk)$-fence in $S'$ the algebraic length of $\indg{\graph{G}}{S'}$ is one.
\qed

\noindent
It remains to prove the case of Theorem~\ref{thm:realsmooth} when there is no absorbing subuniverse in any weak component of $\graph{G}$ of algebraic length one. We choose such a component and call it $H$. By Claim~\ref{claim:abs2} there is an $a\in H$ and $\Fence$ such that $H=\{a\}^{\Fence}$. Since $\{a\}$ is a subuniverse, multiple application of Lemma \ref{lem:neig} (as above) shows that $H$ is a subuniverse as well. If $H\varsubsetneq A$ we are done by the inductive assumption. Therefore $H=A$ and there is no absorbing subuniverse in $\alg{A}$. 

Let $k$ be minimal such that there exists $m$ and $a\in A$ with $\{a\}^{\fence{k}{m}} = A$. This implies that $E^{\circ k}\leq_S A\times A$ is linked and, as there is no absorbing subuniverse in $\alg{A}$, Theorem~\ref{thm:abs} implies that $E^{\circ k} =  A\times A$. In particular the digraph $\graph{G}$ is strongly connected. Choose any $a\in A$ and consider the fence $\fence{k-1}{m'}$ for $m'$ large enough so that $B=\{a\}^{\fence{k-1}{m'}}=\{a\}^{\fence{k-1}{m'+1}}$. The set $B$ is a proper subset of $A$ (by minimality of $k$) and it is a  subuniverse of  $\alg{A}$~(by Lemma \ref{lem:neig} again). It suffices to prove that the smooth part of $\indg{\graph{G}}{B}$~(which is a subuniverse by Lemma~\ref{lem:bothways}) has algebraic length $1$.

\begin{claim}
 The smooth part of $\indg{\graph{G}}{B}$, denoted by $B'$, is non-empty and has algebraic length one.
\end{claim}
\proof
 Note that, by definition of $B$, $B^{\fence{k-1}{1}} = B$. 
 
Let $b$ be an arbitrary element of $B$. As $\graph{G}$ is smooth we can find  $c\in A$ such that $b\xrightarrow{k-1} c$. Since $E^{\circ k} = A\times A$ we get $b\xrightarrow{k} c$. Consider the first element $b_1$ on this path: $b\rightarrow b_1$ and $b_1\in B$ as $b\xrightarrow{k-1}c\xleftarrow{k-1}b_1$. Therefore $b\rightarrow b_1$ in $\indg{\graph{G}}{B}$. We have shown that $B^{\cdot\leftarrow\cdot}\supseteq B$. By Proposition~\ref{prop:basicsmooth} the smooth part of $B$ is non-empty.

To show that $\indg{\graph{G}}{B'}$ has algebraic length one we pick arbitrary $b,b' \in B'$ such that $b\xrightarrow{k-1}b'$ in $\indg{G}{B'}$.
As $E^{\circ k} = A\times A$ we have $b\xrightarrow{k} b'$ in $\graph{G}$.
All the vertices on the path $b\xrightarrow{k} b'$ are in $B$, because $B^{\fence{k-1}{1}} = B$ and $b'$ is in the smooth part of $\indg{G}{B}$.
Since $b,b'$ are in $B'$, the whole path falls in $B'$.  
This gives a path of algebraic length one connecting $b$ to $b$ in $B'$ which proves the claim.
%
\qed

\section{Cyclic terms in Taylor varieties}\label{sect:cyclic}

\noindent
In the final section we prove our second main result -- a characterization of Taylor varieties as the varieties possessing a cyclic term.

\begin{thm}\label{thm:cyclic}
 Let $\variety{V}$ be an idempotent variety generated by a finite algebra $\alg{A}$ then the following are equivalent.
\begin{iteMize}{$\bullet$}
 \item $\variety{V}$ is a Taylor variety;
 \item $\variety{V}$~(equivalently the algebra $\alg{A}$) has a cyclic term;
 \item $\variety{V}$~(equivalently the algebra $\alg{A}$) has a cyclic term of arity $p$, for every prime $p>|A|$.
\end{iteMize}
\end{thm}

\noindent
The proof uses the Absorption Theorem and its corollaries, and Theorem \ref{thm:realsmooth}.
This result is then applied to restate the Algebraic Dichotomy
Conjecture, and to give short proofs of Theorem \ref{thm:wnu} and the dichotomy theorem for undirected graphs \cite{HN90}. 
At the very end of the section we provide more information about possible arities of cyclic terms of a finite algebra.

\subsection{Proof of Theorem \ref{thm:cyclic}}
As every cyclic term is a Taylor term, Theorem~\ref{thm:cyclic} will follow immediately when we prove:

\begin{thm}\label{thm:cyclicreal} 
Let $\alg{A}$ be a finite algebra in a Taylor variety and let $p$ be a prime such that $p > |A|$. Then
$\alg{A}$ has a $p$-ary cyclic term operation.
\end{thm}

\noindent
As in the proofs of partial results \cite{firstcyclic,sdjoin}, the proof of Theorem \ref{thm:cyclicreal} is based on studying cyclic relations:

\begin{defi}
An $n$-ary relation $R$ on a set $A$  is called \emph{cyclic}, if for all $a_0,\dotsc,a_{n-1}\in A$
\begin{equation*}
(a_0, a_1, \dots, a_{n-1}) \in R \ \  \Rightarrow \ \ (a_1, a_2, \dots, a_{n-1}, a_0) \in R.
\end{equation*}
\end{defi}

\noindent The following lemma from \cite{firstcyclic} gives a connection between cyclic operations and cyclic relations. 
\begin{lem}\label{lem:cyclicbysub}
For a finite, idempotent algebra $\alg{A}$ the following are equivalent:
\begin{iteMize}{$\bullet$}
 \item $\alg{A}$ has a $k$-ary cyclic term operation;
 \item every nonempty cyclic subalgebra of $\alg{A}^k$ contains a constant tuple.
\end{iteMize}
\end{lem}

\proof 
Assume first that $\alg{A}$ has a $k$-ary cyclic term operation $t$ and consider an arbitrary tuple $\tuple{a} = (a_0,a_1, \dots, a_{k-1})$ in a cyclic subalgebra $\alg{R}$ of $\alg{A}^k$. We denote by $\sigma(\tuple{a})$, $\sigma^2(\tuple{a})$, \dots, $\sigma^{k-1}(\tuple{a})$ the cyclic shifts of $\tuple{a}$, that is $\sigma(\tuple{a}) = (a_1, a_2, \dots, a_{k-1},a_0)$, $\sigma^2(\tuple{a}) = (a_2, a_3, \dots, a_{k-1},a_0,a_1)$, \dots, $\sigma^{k-1}(\tuple{a}) = (a_{k-1},a_0, a_1, \dots, a_{k-2})$. 
As $R$ is cyclic, all these shifts belong to $R$. By applying $t$ to the tuples $\tuple{a}$, $\sigma(\tuple{a})$, \dots, $\sigma^{k-1}(\tuple{a})$ coordinatewise we get the tuple
$$
(t(a_0,a_1, \dots, a_{k-1}), t(a_1, a_2, \dots, a_{k-1},a_0), \dots, t(a_{k-1},a_0, a_1, \dots, a_{k-2})),
$$
which belongs to $R$, since $R$ is a subuniverse of $\alg{A}^k$. But $t$ is a cyclic operation, therefore this tuple is constant.

To prove the converse implication, we assume that every nonempty cyclic subalgebra of $\alg{A}^k$ contains a constant tuple.
For a $k$-ary operation $t\in\Clo(\alg{A})$ we define $S(t)\subseteq A^k$ to be the set of all $\tuple{a}  \in A^k$ such that $t(\tuple{a})=t(\sigma(\tuple{a}))=\dots=t(\sigma^{k-1}(\tuple{a}))$. 
Let $t$ be such that $|S(t)|$ is maximal. 

If $S(t) = A^k$, then the term operation $t$ is cyclic and we are done. Assume the contrary, that is, there exists a tuple $\tuple{a} \in A^k$ such that $t(\tuple{a})=t(\sigma(\tuple{a}))=\dots=t(\sigma^{k-1}(\tuple{a}))$ fails. 
Consider the tuple $\tuple{b} = (b_0, b_1, \dots, b_{k-1})$ defined by $b_i=t(\sigma^{i}(\tuple{a}))$, $0 \leq i < k$, and 
let $B = \{\tuple{b}, \sigma(\tuple{b}), \dots, \sigma^{k-1}(\tuple{b})\}$. 

We claim that the subalgebra $\alg{C} = \Sg{\alg{A}^k}{B}$ of $\alg{A}^k$ is cyclic. Indeed, every tuple $\tuple{c} \in C$ can be written as
$\tuple{c} = s(\tuple{b}, \sigma(\tuple{b}), \dots, \sigma^{k-1}(\tuple{b}))$ for some term $s$. Then the element $s(\sigma(\tuple{b}), \sigma^2(\tuple{b}), \dots, \sigma^{k-1}(\tuple{b}), \tuple{b})$ of $\alg{C}$ is equal to $\sigma(\tuple{c})$.

According to our assumption, the algebra $\alg{C}$ contains a constant tuple. It follows that there exists a $k$-ary term $s \in \Clo(\alg{A})$ such that $\tuple{b} \in S(s)$. Now consider the term $r$ defined by
$$
r(x_0, x_1, \dots, x_{k-1}) = s(t(x_0, x_1, \dots, x_{k-1}), t(x_1, \dots, x_{k-1},x_0), \dots, t(x_{k-1},x_0, x_1, \dots, x_{k-2})).
$$ 
We claim that $S(t)\subseteq S(r)$, but also that $\tuple{a} \in S(r)$. This would clearly be a contradiction with the maximality of $|S(t)|$.
Let $\tuple{x}\in S(t)$. Then 
$$
r(\sigma^i(\tuple{x}))=s(t(\sigma^i(\tuple{x})),t(\sigma^{i+1}({\tuple{x}})),\dots,t(\sigma^{i-1}({\tuple{x}})))=s(t(\tuple{x}),t(\tuple{x}),\dots,t(\tuple{x}))=t(\tuple{x})
$$ 
for all $i$, so $\tuple{x}\in S(r)$. 
On the other hand, $$
r(\sigma^i(\tuple{a}))=s(t(\sigma^i(\tuple{a})),t(\sigma^{i+1}(\tuple{a})),\dots,t(\sigma^{i-1}(\tuple{a})))=s(b_i,b_{i+1},\dots,b_{i-1})=s(\sigma^i(\tuple{b})),$$
 which is constant for all $i$ by the choice of $s$. Therefore $\tuple{a}\in S(r)$ and the contradiction is established.
\qed

\noindent
For the rest of the proof of Theorem \ref{thm:cyclicreal} we fix a prime number $p$, we fix a Taylor variety $\variety{V}$ and we consider a minimal counterexample to the theorem with respect to the size of $A$.
Thus $\alg{A}$ is a finite algebra in $\variety{V}$, $p > |A|$, and for all $\alg{B} \in \variety{V}$ with $|B| < |A|$, $\alg{B}$ has a cyclic term of arity $p$, i.e., by Lemma~\ref{lem:cyclicbysub}, every nonempty cyclic subuniverse of $\alg{B}^p$ contains a constant tuple. 

An easy reduction proving the following claim can also be found in~\cite{firstcyclic}.

\begin{claim} \label{clm:simple}
$\alg{A}$ is simple.
\end{claim}

\proof
Suppose that $\alg{A}$ is not simple, and $\alpha$ is a nontrivial congruence of $\alg{A}$.

To apply Lemma~\ref{lem:cyclicbysub} we focus on an arbitrary cyclic subalgebra $\alg{R}$ of $\alg{A}^p$. Our first objective is to find a tuple in $\alg{R}$ with all elements congruent to each other modulo $\alpha$. Let us choose any tuple $(a_0,\dotsc,a_{k-1})\in R$ and let $c(x_0,\dotsc,x_{k-1})$ be the operation of $\alg{A}$ which gives rise to the cyclic operation of $\alg{A}/\alpha$ (such an operation exists from the minimality assumption). Therefore $c(a_0,\dotsc,a_{k-1}), c(a_1,\dotsc,a_{k-1},a_0),\dotsc$ all lie in one congruence block of $\alpha$ as the results of these evaluations are equal in $\alg{A}/\alpha$. Now we apply the term $c(x_0,\dotsc,x_{k-1})$ in $\alg{R}$ to $(a_0,\dotsc,a_{k-1}), (a_1,\dotsc,a_{k-1},a_0),\dotsc$ and obtain the tuple $(c(a_0,\dotsc,a_{k-1}), c(a_1,\dotsc,a_{k-1},a_0),\dotsc)$ in $\alg{R}$ with all the coordinates in the same congruence block.

Let $C$ be a congruence block of $\alpha$ such that $C^p\cap R\neq\emptyset$. It is easy to see that in such a case $C^p\cap R$ is a (nonempty) cyclic subuniverse of $\alg{C}^p$. As the block $C$ has a cyclic operation of arity $p$ then, again by Lemma~\ref{lem:cyclicbysub}, we obtain a constant in $C^p\cap R$ and the claim is proved.
\qed

\noindent
From Lemma~\ref{lem:cyclicbysub} it follows that there exists a cyclic subalgebra $\alg{R}$ of $\alg{A}^p$ containing no constant tuple.
We fix such a subalgebra $\alg{R}$.
Let $\alg{R}_k$, $k=1,2, \dots, p$, denote the projection of $\alg{R}$ to the first $k$ coordinates, that is
$$
R_k = \{ (a_0, a_1, \dots, a_{k-1}) : (a_0,\dotsc,a_{p-1})  \in R \}.
$$
Note that, from the cyclicity of $R$, it follows that for any $i$ we have
$$
R_k = \{ (a_i, a_{i+1}, \dots, a_{i+k-1}) : (a_0,\dotsc,a_{p-1}) \in R \},
$$
where indices are computed modulo $p$. In the next claim we show that $R$ is subdirect in $\alg{A}^p$.

\begin{claim} \label{cl:subdir}
$R_1 = A$.
\end{claim}

\proof
The projection of $R$ to any coordinate is a subalgebra of $\alg{A}$. From the cyclicity of $R$ it follows that all the projections  are equal, say to $B$. The set $B$ is a subuniverse of $\alg{A}$ and if it is a proper subset of $A$, then $R \leq_S \alg{B}^p$ contains a constant tuple by the minimality assumption, a contradiction.
\qed

\noindent
We will prove the following two claims by induction on $n=1,2, \dots, p$.  Note that for $n=1$ both claims are valid and that property (P1) for $n=p$ contradicts the absence of a constant tuple in $R$.
\begin{enumerate}[(P1)]
 \item There exists $\alg{I} \minabs \alg{A}$ such that $\alg{I}^n \minabs \alg{R}_n$.
 \item If $\alg{I}_1, \dots, \alg{I}_n \minabs \alg{A}$ and $(I_1 \times \dots\times I_n) \cap R_n \neq \emptyset$, then
   $\alg{I}_1 \times \dots \times \alg{I}_n \minabs R_{n}$.
\end{enumerate}
We assume that both (P1) and (P2) hold for some $n \in \{1, \dots, p-1\}$ and we aim to prove these properties for $n+1$.
We fix 
$\alg{I} \minabs \alg{A}$ such that $\alg{I}^n \minabs \alg{R}_n$ guaranteed by (P1). Let 
$$
S = \{ ( (a_0, \dots, a_{n-1}), a_{n}) : (a_0, \dots, a_n) \in R_{n+1} \}
$$
and let $\alg{S}$ denote the subalgebra of $\alg{A}^{n+1}$ with universe $S$. Thus $\alg{S}$ is basically $\alg{R}_{n+1}$, but we look at it as a (subdirect) product of two algebras $\alg{R}_n$ and $\alg{A}$: $S \leq_S \alg{R}_{n} \times \alg{A}$.

The aim of the next few claims is to show that $S$ is linked. 
First we show, that it is enough to have a ``fork''.

\begin{claim} \label{cl:fork}
 If there exist $\tuple{a} \in \alg{R}_n$ and $b,b' \in A$, $b \neq b'$ such that $(\tuple{a},b),(\tuple{a},b') \in S$, then $S$ is linked.
\end{claim}

\proof
Let $k = |A|$. We define a binary relation $\sim$ on $A$ by putting $b \sim b'$ if and only if there exist tuples $\tuple{a}^1, \dots, \tuple {a}^k \in R_n$ and  elements $b=c_0, c_1, \dots, c_k=b' \in A$ such that for every $i \in \{1,2, \dots, k\} $ we have
$$ (\tuple{a}^i,c_{i-1}), (\tuple{a}^i, c_i) \in S.$$
The relation $\sim$ is clearly reflexive and symmetric. It is also transitive as we have chosen $k$ big enough. 
It follows immediately from the definition that $\sim$ is a subuniverse of $\alg{A}^2$.

Therefore $\sim$ is a congruence of $\alg{A}$.  Moreover, from the assumption of the claim it follows that it is not the smallest congruence~(as $b\sim b'$ for $b\neq b'$). Since, by Claim \ref{clm:simple}, $\alg{A}$ is simple, then $\sim$ is the full relation on $A$ and therefore $S$ is linked.
\qed

\noindent
The next claim shows that $S$ is linked in case that $\alg{A}$ has no proper absorbing subuniverse.

\begin{claim}\label{cl:binlinked}
 If $I = A$ then $S$ is linked.
\end{claim}

\proof
From (P1) we have $R_n = A^n$.  If there are $(a_0,\dotsc,a_{p-1}),(b_0,\dotsc,b_{p-1}) \in R$ such that $a_i \neq b_i$ for some $i$ and $a_0 =b_0$, $a_1=b_1$, \dots, $a_{i-1} = b_{i-1}$, $a_{i+1} = b_{i+1}$, \dots, $a_{n-1}=b_{n-1}$, then, by cyclically shifting these tuples, we obtain tuples $(a'_0, a'_1, \dots, a'_{p-1})$ and $(b'_0, b'_1, \dots, b'_{p-1})$ such that $a'_0 = b'_0$, \dots, $a'_{n-1}=b'_{n-1}$, and $a_n \neq b_n$. Then Claim~\ref{cl:fork} proves that $S$ is linked.

In the other case, tuples in $R$ are determined by the first $n$ projections, thus $|R| = |R_n| = |A|^n$. Consider the mapping $\sigma: R \rightarrow R$ sending a tuple $(a_0, \dots, a_{p-1}) \in R$ to its cyclic shift $(a_1, \dots, a_{p-1},a_0) \in R$. Clearly, $\sigma$ is a permutation of $R$ satisfying $\sigma^{p} = \mathrm{id}$. Now $p$ is a prime number and $|R|=|A|^n$ is not divisible by $p$ (as $p > |A|$), therefore $\sigma$ has a fixed point, that is, a constant tuple. A contradiction.
\qed

\noindent
The harder case is when $I \neq A$. 
We need two more auxiliary claims.

\begin{claim} \label{cl:j}
 If $I \neq A$ then there exists $\alg{J} \minabs \alg{A}$ such that  $I \neq J$ and
$(I^n \times J) \cap R_{n+1} \neq \emptyset$.
\end{claim}

\proof
Observe that $I^p \cap R$ is a cyclic subuniverse of $\alg{I}^p$ without a constant tuple. Therefore, by minimality, the intersection $I^p \cap R$ is empty.
On the other hand $I^n \cap R_n \neq \emptyset$ by (P1), so that there exists a greatest number
$k$, $n \leq k < p$,  such that $(I^k\times A^{p-k}) \cap R$ is nonempty. Consider the set
$$
X = \{a: (a_0, \dots, a_{k-1},a) \in R_{k+1}, \  \ a_0, \dots, a_{k-1} \in I\}.
$$
It is easy to check that $X$ is an absorbing subuniverse of $\alg{A}$. As $I^{k+1} \cap R_{k+1}$ is empty, $X$ is disjoint from $I$. Let $J$ be a minimal absorbing subuniverse of $\alg{X}$. 
We have $J \minabs \alg{A}$ (as $J \minabs \alg{X} \absorb \alg{A}$), $I \neq J$ and $(I^k \times J) \cap R_{k+1} \neq \emptyset$.
We take a tuple in $R$ whose projection to the first $(k+1)$ coordinates lies in $I^k \times J$, and shift it $(k-n)$ times to the left (recall that $k-n \geq 0$). This tuple shows that $(I^n \times J) \cap R_{n+1}$ is nonempty.
\qed

\noindent
Similarly we can show that there exists a minimal absorbing subalgebra $\alg{J}'$ of $\alg{A}$ distinct from $I$ such that $(J' \times I^n) \cap R_{n+1}$ is nonempty.

We consider the following two subsets of $A \times A$.
\begin{eqnarray*}
F &=& \{ (a,b) :  \exists \ (a,c_1, \dots, c_{n-1},b) \in R_{n+1} \}  \\
E &=& \{ (a,b) : \exists \  (a,c_1, \dots, c_{n-1},b) \in R_{n+1}\  \mbox{ and } \ \forall i \ c_i\in I\} 
\end{eqnarray*}
Let $V_1$ and $V_2$ denote the projections of $E$ to the first and the second coordinate, so that $E \subseteq_S V_1 \times V_2$.

\begin{claim}
 $E$ is a subuniverse of $\alg{A}^2$, is linked and subdirect in $V_1\times V_2$  and $V_1, V_2 \absorb \alg{A}$.
\end{claim}

\proof
It is straightforward to check that $E$ and $F$ are subuniverses of $\alg{A}^2$, that $\alg{E} \absorb \alg{F}$ and that $\alg{V}_1, \alg{V}_2 \absorb \alg{A}$, where $\alg{E}, \alg{F}$ denote the subalgebras of $\alg{A}^2$ with universes $E,F$ and $\alg{V}_1, \alg{V}_2$ denote the subalgebras of $\alg{A}$ with universes $V_1, V_2$.
From Claim \ref{cl:subdir} we know that $F \leq_S A \times A$. 

Similarly as in the proof of Claim \ref{cl:fork} we will show that $F$ is linked. Let $k = |A|$ and let us define a congruence $\sim$ on $\alg{A}$ by putting
$b \sim b'$ if and only if there are  $a_1, a_2, \dots, a_{k}, b=b_0, b_1, \dots, b_k = b' \in A$ such that for all $i \in \{1,2, \dots, k\}$
$$
(a_i,b_{i-1}), (a_i,b_i) \in F.
$$
The proof that $\sim$ is a congruence follows exactly as in Claim~\ref{cl:fork}.

Take an arbitrary tuple $(a_0,\dotsc,a_{p-1}) \in R$. As $p$ is greater than $|A|$ we can find indices $i \neq j$ such that $a_i = a_j$.
There exists $k$ such that $a_{i+kn} \neq a_{j+kn}$~(indices computed modulo $p$), otherwise (as $p$ is a prime number) the tuple would be constant. 
It follows that there exist $i',j'$ such that $a_{i'} = a_{j'}$ and $a_{i'+n} \neq a_{j'+n}$. The pairs $(a_{i'},a_{i'+n})$ and $(a_{j'},a_{j'+n})$ are in $F$ (by shifting $(a_0,\dotsc,a_{p-1})$), therefore $\sim$ is not the smallest congruence. Since $\alg{A}$ is simple, $\sim$ is the full congruence on $\alg{A}$, thus $F$ is linked. By Proposition \ref{prop:abscor}.(i), $E$ is linked as well.
\qed

\noindent
Now we can finally show that $S$ is linked.

\begin{claim} \label{cl_con}
$S$ is linked.
\end{claim}
\proof
From Claim \ref{cl:j} and the remark following it we know that $(a,b'),(a',b) \in E$ for some $a,b \in I, a' \in J', b' \in J$, $J,J' \minabs \alg{A}$, $I \neq J$, $I \neq J'$.
As $E$ is linked, we can find  elements $a = c_0, c_1, \dots, c_{2i} = a'$ such that
$c_0, c_2, \dots, c_{2i} \in V_1$, $c_1, c_3, \dots, c_{2i-1} \in V_2$ and $(c_{2j},c_{2j+1}), (c_{2j+2},c_{2j+1}) \in E$ for all $j=0,1, \dots, i-1$. 
By Proposition \ref{prop:abscor}.(v) (used for $\alg{E} \leq_S \alg{V}_1 \times \alg{V}_2$) 
we can assume that all the elements $c_0, \dots c_{2i}$ lie in minimal absorbing subuniverses of $\alg{V}_1$ or $\alg{V}_2$ (which are also minimal absorbing subuniverses of $\alg{A}$, since $V_1, V_2 \absorb \alg{A}$).
It follows that
there exist $w \in W \minabs \alg{V}_1$ and $u \in U \minabs \alg{V}_2, v \in V \minabs \alg{V}_2$ such that $(w,u),(w,v) \in E$, $U \neq V$.
Therefore there exist $a_1, \dots, a_{n-1}, a_1', \dots, a'_{n-1} \in I$ such that $(w,a_1, \dots, a_{n-1},u), (w,a_1',\dots, a_{n-1}',v) \in R_{n+1}$.

From the induction hypotheses (P2) we know that $W \times I^{n-1} \minabs {R}_n$. Also $V \minabs \alg{A}$ and $((W \times I^{n-1}) \times V) \cap S \neq \emptyset$. 
By Proposition \ref{prop:abscor}.(ii), $((W \times I^{n-1}) \times V) \cap S \leq_S (W \times I^{n-1}) \times V$. 
In particular, there exists $v' \in V$  such that $(w,a_1, \dots, a_{n-1},v') \in R_{n+1}$.
Now recall that $(w,a_1, \dots, a_{n-1},u) \in R_{n+1}$ and observe that $u$ and $v'$ are distinct, since they lie in different minimal absorbing subuniverses. Then
$S$ is linked by Claim \ref{cl:fork}.
\qed

\noindent
We are ready to prove (P2) for $n+1$. 

\begin{claim} \label{cl:ptwo}
 (P2) holds for $n+1$.
\end{claim}
\proof
Let $\alg{I}_1, \dots, \alg{I}_{n+1}$ be absorbing subalgebras of $\alg{A}$ such that $(I_1 \times \dots \times I_{n+1}) \cap R_{n+1} \neq \emptyset$.
Now $S$ is a linked subdirect subuniverse of $\alg{R}_n \times \alg{A}$, $I_1 \times \dots \times I_n$ is a minimal absorbing subuniverse of $\alg{R}_n$ (from the induction hypotheses (P2)), $I_{n+1} \minabs \alg{A}$ and $((I_1 \times \dots \times I_n) \times I_{n+1}) \cap S \neq \emptyset$. 
By Proposition \ref{prop:abscor}.(iii), $(I_1 \times \dots \times I_n) \times I_{n+1}$ is a minimal absorbing subuniverse of $\alg{S}$ and thus
$I_1 \times \dots \times I_{n+1}$ is a minimal absorbing subuniverse of $\alg{R}_{n+1}$.
\qed

\noindent
To prove (P1) for $n+1$ we define a digraph on the vertex set $R_n$ by putting 
$$
 ((a_0, \dots, a_{n-1}), (a_1, \dots, a_n))\in H
$$
whenever $(a_0,\dotsc,a_{n})\in R_{n+1}$. We want to apply Theorem \ref{thm:realsmooth} to obtain a loop of the digraph $\relstr{G} =(R_n,H)$ in a minimal absorbing subuniverse of $\alg{R}_n$. 

Observe that $H$ is a subuniverse of $\alg{R}_n^2$. Next we show that $I^n$ is contained in a weak component of $\relstr{G}$.

\begin{claim}
  Any two elements of $I^n$ are in the same weak component of the digraph $\relstr{G}$.
\end{claim}

\proof
The set
$X = \{x: (a_0, \dots, a_{n-1},x) \in R_{n+1}, \ \ a_0, \dots, a_{n-1} \in I\}$ is an absorbing subuniverse of $\alg{A}$. Let
$X_0$ be a minimal absorbing subuniverse of the algebra $\alg{X}$ with universe $X$. We have found $X_0 \absorb \alg{A}$ such that
$(I^n \times X_0) \cap R_{n+1} \neq \emptyset$. Similarly we can find $X_1, X_2, \dots, X_{n-1}$ such that
$(I^{n-i} \times X_0 \times X_1 \times \dots \times X_i) \cap R_{n+1} \neq \emptyset$ for all $i=0,1, \dots, n-1$.
From (P2) for $n+1$ (Claim \ref{cl:ptwo}) it follows that $I^{n-i} \times X_0 \times X_1 \times \dots \times X_i \subseteq R_{n+1}$ for
all $i$. Now choose arbitrary elements $x_i \in X_i$ and take any tuple $(b_0, \dots, b_{n-1}) \in I^n$.
Since, for all $i =0,1,\dots, n-1$, the tuple $(b_i, \dots, b_{n-1},x_0,x_1, \dots, x_i)$ belongs to $R_{n+1}$, 
the vertices $(b_i, \dots, b_{n-1},x_0, \dots, x_{i-1})$ and $(b_{i+1}, \dots, b_{n-1},x_0, \dots, x_i)$ are in the same weak component of $\relstr{G}$.
Therefore the vertex $(b_0, \dots, b_{n-1})$, which was an arbitrarily chosen vertex in $I^n$, is in the same weak component as the vertex $(x_0, \dots, x_{n-1})$.
\qed

\noindent
The last assumption of Theorem \ref{thm:realsmooth} is proved in the next claim.

\begin{claim}
The weak component of $\relstr{G}$ containing $I^n$ has algebraic length $1$.
\end{claim}

\proof
Let $b\in I$ be arbitrary. As $E$ is linked, $b \in V_1$ can be $E$-linked to $b \in V_2$, i.e.
there exist $b=c_0, c_1, \dots, c_{2i}$ such that $(c_{2j},c_{2j+1}),\,(c_{2j+2},c_{2j+1})\in E$ for all $j=0, \dots, i-1$ and $(c_{2i},b) \in E$.
By Proposition~\ref{prop:abscor}.(v) we can assume that these elements lie in minimal absorbing subuniverses of $\alg{A}$.
Property (P2) for $n+1$ (Claim \ref{cl:ptwo}) proves that $(c_{2j}, b,\dots, b, c_{2j+1}), (c_{2j+2},b, \dots, b,c_{2j+1})\in R_{n+1}$ for all $j=0, \dots, i-1$ and $(c_{2i},b, \dots, b,b) \in R_{n+1}$. This gives rise to a $(1,j)$-fence connecting, in $\graph{G}$, the tuple $(c_0=b,\dotsc,b)$ to the tuple $(c_{2i},b,\dots,b)$. As $((c_{2i},b, \dots,b), (b,\dotsc,b))\in H$ we showed that the algebraic length of the weak component containing $I^n$ is one.
\qed

\noindent
By Theorem \ref{thm:realsmooth} there exists a loop inside a minimal absorbing subuniverse $K$ of $\alg{R}_{n}$. Since the projection $J$ of $K$ to the first coordinate is a minimal absorbing subuniverse of $\alg{A}$, we actually get an element $a \in J \minabs \alg{A}$ such that $(a, \dots, a) \in R_{n+1}$. Now (P1) follows from (P2) and the proof of Theorem \ref{thm:cyclicreal} is concluded.

\subsection{Consequences of Theorem~\ref{thm:cyclic}}
First we restate the hardness criterion in Theorem~\ref{thm:bjk} and
the Algebraic Dichotomy Conjecture of Bulatov, Jeavons and
Krokhin. These statements are equivalent to the original ones by
Theorem~\ref{thm:cyclic} and Lemma~\ref{lem:cyclicbysub}.

\begin{thm} \label{thm:hardpp}
Let $\relstr{A}$ be a core relational structure and let $p$ be a prime
number greater than the size of the universe of $\relstr{A}$. If there
exists a nonempty positively primitively defined cyclic $p$-ary relation
without a constant tuple then $\CSP(\relstr{A})$ is NP-complete.
\qed
\end{thm}

\begin{algdich}
 Let $\relstr{A}$ be a a core relational structure. Let $p$ be a prime
 number greater than the size of the universe of $\relstr{A}$. 
If every nonempty positively primitively defined cyclic $p$-ary relation has a constant tuple then $\CSP(\relstr{A})$ is solvable in polynomial time. Otherwise it is NP-complete.
\end{algdich}

\noindent 
As a second consequence we reprove the dichotomy theorem of Hell and Ne\v set\v ril~\cite{HN90}. 
It follows immediately from the Smooth Theorem from Section~\ref{sect:smoothproof}, but the following proof is an elegant way of presenting it.

\begin{cor}[Hell and Ne\v set\v ril~\cite{HN90}]
 Let $\graph{G}$ be an undirected graph without loops. If $\graph{G}$ is bipartite then $\CSP(\graph{G})$ is solvable in polynomial time. Otherwise it is NP-complete.
\end{cor}

\proof
 Without loss of generality we can assume that $\graph{G}$ is a core. If the graph $\graph{G}$ is bipartite then it is a single edge and $\CSP(\graph{G})$ is solvable in polynomial time. Assume now that $\graph{G}$ is not bipartite --- therefore there exists a cycle $a\xrightarrow{2k+1}a$ of odd length in $\graph{G}$. As vertex $a$ is in a $2$-cycle~(i.e. an undirected edge) we can find a path $a\xrightarrow{i(2k+1)+j2} a$ for any non-negative numbers $i$ and $j$. Thus, for any number $l\geq 2k$ we have $a\xrightarrow{l}a$. Let $p$ be any prime greater than $\max\{2k,|A|\}$ and $t$ be any $p$-ary polymorphism of $\graph{G}$. Let $a=a_0\rightarrow a_1\rightarrow\dotsb\rightarrow a_{p-1}\rightarrow a$. Then 
\begin{equation*}
 t(a_0,\dotsc,a_{p-1})\rightarrow t(a_1,\dotsc, a_{p-1},a_0)
\end{equation*}
and, if $t$ were a cyclic operation we would have
\begin{equation*}
 t(a_0,\dotsc,a_{p-1}) = t(a_1,\dotsc, a_{p-1},a_0)
\end{equation*}
which implies a loop in $\graph{G}$. This contradiction shows that
$\graph{G}$ has no cyclic polymorphism for some prime greater than the
size of the vertex set which, by Theorem~\ref{thm:cyclic}, implies
that the associated variety is not Taylor and therefore, by
Theorem~\ref{thm:bjk}, $\CSP(\graph{G})$ is NP-complete.

Equivalently one can consider the relation
$$
R = \{ (a_0, \dots, a_{p-1}) : a_0 \rightarrow a_1 \rightarrow a_2
\rightarrow \dots \rightarrow a_{p-1} \rightarrow a_0 \},
$$
where $p$ is chosen as above. It is easy to see that $R$ is a cyclic, 
positively primitively defined nonempty relation without a constant tuple and
therefore $\CSP(\graph{G})$ is NP-complete by Theorem~\ref{thm:hardpp}.
\qed

\noindent
Finally, we observe that the weak near-unanimity characterization of
Taylor varieties (Theorem~\ref{thm:wnu}) is  a consequence of
Theorem~\ref{thm:cyclic}:

\begin{cor}[Maroti and McKenzie~\cite{MM}]
 For a locally finite idempotent variety $\variety{V}$ the following are equivalent.
\begin{iteMize}{$\bullet$}
 \item $\variety{V}$ is a Taylor variety;
 \item $\variety{V}$ has a weak near-unanimity term.
\end{iteMize}
\end{cor}

\proof
In the case that $\variety{V}$ is finitely generated, the theorem is an immediate consequence of Theorem \ref{thm:cyclic}.
In the general case the proof can be done by a standard universal algebraic argument --- we apply Theorem \ref{thm:cyclic} to the \emph{free algebra} on two generators.
\qed
As opposed to the previous theorem the assumption in Theorem \ref{thm:cyclic} that $\variety{V}$ is finitely generated cannot be relaxed to locally finite ~\cite{firstcyclic}.

It was observed by Matt Valeriote \cite{mattcommun} that Sigger's characterization of Taylor varieties \cite{sig} is also an easy corollary of Theorem \ref{thm:cyclic}. The proof will appear elsewhere.

\subsection{Arities of cyclic terms}
Let $\alg{A}$ be a finite algebra and let $C(\alg{A})$ be the set of arities of cyclic operations of $\alg{A}$ i.e.:
\begin{equation*}
C(\alg{A}) = \{n : \alg{A} \mbox{ has a cyclic term of arity $n$} \}.
\end{equation*}

\noindent The following simple proposition was proved in~\cite{firstcyclic}.
\begin{prop}[\cite{firstcyclic}]
Let $\alg{A}$ be a finite algebra let $m,n$ be natural numbers. Then the following are equivalent.
\begin{enumerate}[\em(i)]
\item $m,n\in C(\alg{A})$;
\item $mn\in C(\alg{A})$.
\end{enumerate}
\end{prop}

\noindent 
This implies that $C(\alg{A})$ is fully determined by its prime elements. 
There are algebras in Taylor varieties with no cyclic terms of arities smaller than their size~\cite{firstcyclic}. 
However the following simple lemma provides, under special circumstances, additional elements in $C(\alg{A})$.
Its proof follows the lines of the proof of Claim~\ref{clm:simple}.

\begin{lem}
  Let $\alg{A}$ be a finite, idempotent algebra and $\alpha$ be a congruence of $\alg{A}$. If $\alg{A}/\alpha$ and every $\alpha$-block in $A$ have cyclic operation of arity $k$ then so does $\alg{A}$. \qed
\end{lem}

\noindent This leads to the following observation.

\begin{cor}
 Let $\alg{A}$ be a finite, idempotent algebra in Taylor variety. Let $0_A = \alpha_0 \subseteq \dotsb \subseteq \alpha_n = 1_A$ be an increasing sequence of congruences on $\alg{A}$. If $p$ is a prime number such that, for every $i\geq 1$, every class of $\alpha_i$ splits into less than $p$ classes of $\alpha_{i-1}$ then $\alg{A}$ has a $p$-ary cyclic term. \qed
\end{cor}

\newcommand{\etalchar}[1]{$^{#1}$}
\def\cprime{$'$}


\end{document}